\documentclass[usenatbib]{mn2e}
\usepackage{mathtools}
\usepackage{graphicx}
\usepackage{epstopdf}
\usepackage{natbib}
\usepackage{verbatim}

\pdfoutput=1

\renewcommand{\thefootnote}{\fnsymbol{footnote}}

\newcommand\araa{ARA\&A}
\newcommand\apj{ApJ}
\newcommand\apjl{ApJ}
\newcommand\apjs{ApJS}
\newcommand\aap{A\&A}
\newcommand\mnras{MNRAS}
\newcommand\nat{Nature}

\title[Moving Mesh Cosmology: Gas Disks]{Moving Mesh Cosmology: Properties of Gas Disks}

\author[Torrey et al.]
       {\parbox{18cm}{Paul~Torrey$^{1}$\footnotemark[1], Mark
           Vogelsberger$^{1}$, Debora Sijacki${^1}$\footnotemark[2], Volker
           Springel$^{2,3}$, and Lars Hernquist$^{1}$}\vspace{0.3cm}\\ 
         $^1$ Harvard-Smithsonian Center for Astrophysics, 60 Garden Street,
         Cambridge, MA, 02138, USA\\ 
         $^2$ Heidelberg Institute for Theoretical Studies, Schloss-Wolfsbrunnenweg 35, 69118 Heidelberg, Germany\\
         $^3$ Zentrum f\"{u}r Astronomie der Universit\"{a}t Heidelberg, ARI,
         M\"onchhofstr. 12-14, 69120 Heidelberg, Germany\\}

\begin{document}

\maketitle

\begin{abstract}

We compare the structural properties of galaxies formed in cosmological
simulations using the smoothed particle hydrodynamics (SPH) code {\small
  GADGET} with those using the moving-mesh code {\small AREPO}. Both codes
employ identical gravity solvers and the same sub-resolution physics but use
very different methods to track the hydrodynamic evolution of gas. This
permits us to isolate the effects of the hydro solver on the formation and
evolution of galactic gas disks in {\small GADGET} and {\small AREPO}
haloes with comparable numerical resolution. In a
matching sample of {\small GADGET} and {\small AREPO} haloes we fit simulated
gas disks with exponential profiles. We find that the cold gas disks formed
using the moving mesh approach have systematically larger disk scale lengths
and higher specific angular momenta than their {\small GADGET} counterparts
across a wide range in halo masses. For low mass galaxies differences between
the properties of the simulated galaxy disks are caused by an insufficient
number of resolution elements which lead to the artificial angular momentum
transfer in our SPH calculation. We however find that galactic disks formed in
massive halos, resolved with $\ge 10^6$ particles/cells, are still
systematically smaller in the {\small GADGET} run by a factor of $\sim 2$. The
reason for this is twofold: {\it i)} the excessive heating of haloes close to
the cooling radius due to spurious dissipation of the subsonic turbulence in
{\small GADGET} reduces the supply of gas which can cool and settle onto the
central disk; {\it ii)} the efficient delivery of low angular momentum gaseous
blobs to the bottom of the potential well results in the centrally
concentrated gas disks in {\small GADGET} simulation. While this large
population of gaseous blobs in {\small GADGET} originates from the filaments
which are pressure confined and fragment due to the SPH surface tension while
infalling into hot halo atmospheres, it is essentially absent in the moving mesh
calculation, clearly indicating numerical rather than physical origin of the
blob material. 
\end{abstract}


\section{Introduction}
\renewcommand{\thefootnote}{\fnsymbol{footnote}}
\footnotetext[1]{E-mail: ptorrey@cfa.harvard.edu}
\footnotetext[2]{Hubble Fellow.}

A primary goal of cosmological simulations is to self-consistently reproduce
the variety of galaxy morphologies observed in the local Universe. While the
formation of dark matter haloes via gravitational collapse has been simulated
in great detail using N-body simulations~\citep[e.g.,][]{Millennium,Millennium2,Fosalba2008,Teyssier2009,Bolshoi}, 
modeling the evolution of the luminous components of galaxies has lagged
behind due to the intrinsic complexity of gas dynamics and star
formation.  Early efforts to incorporate baryonic processes into cosmological
simulations accounted for gas cooling, but did not include star formation or
related feedback effects.  These studies found that cooling gas accreted into
dark matter haloes would quickly lose angular momentum and fall to the centre
of the potential~\citep{NavarroBenz1991, KatzGunn91, NavarroWhite94}.  The
forming objects had disk-like morphologies, but with low specific angular
momenta, and with most of the gas residing in a central spheroid rather than a
rotationally supported disk, unlike most observed late-type galaxies.  It was
argued that the efficient angular momentum loss was largely a consequence of
the early collapse and formation of proto-galactic clouds which were able to
efficiently transfer their angular momentum to the dark matter haloes by
dynamical and hydrodynamical friction during merger and accretion events.

A large number of subsequent disk formation studies have attempted to diagnose
and fix this so-called ``angular momentum catastrophe".  Most proposed
solutions are centered around preventing the gas from cooling and forming
stars too efficiently at high redshift.  In a simple experiment,
\citet{WeilEkeEfstathiou98} showed that if gas cooling is prevented until $z = 
1$, stellar disks could form with specific angular momenta consistent with
observed spiral galaxies.  The two widely advocated mechanisms to mitigate gas
over-cooling are heating by an ultraviolet (UV) radiation field and feedback
associated with star formation. The UV background~\citep{Quinn96UVB, 
NavarroSteinmetz97UVB, Hoeft06} has been shown to inhibit the accretion of
cold gas by low mass haloes, but does not appear to provide a full solution to
the angular momentum problem.  Star formation with associated feedback has also
been identified as a heating mechanism which might prevent early
collapse~\citep[e.g.][]{ThackerCouchman2000,ThackerCouchman2001,
MallerDekel2002,Abadi2003, Robertson2004, Okamoto2005,Scannapieco2008} or
efficiently remove low angular momentum
material~\citep[e.g.][]{Governato2009,Guedes2011} which allows less centrally
concentrated disks to form. Although strong feedback can improve galactic
disk formation, it is not immediately clear that this is the only
solution to the angular momentum problem or if other numerical artifacts
remain adversely affecting the formation of rotationally supported
galaxies.

There are some well studied issues with the standard density
formulation of SPH  -- which is the most commonly employed SPH formulation
for cosmological simulation codes -- that can cause spurious angular momentum
transfer from gas disks.  \citet{Okamoto2003} showed that gas disks embedded
in a diffuse hot halo would systematically loose angular momentum due to
spurious hydrodynamical torques, an effect that is particularly severe at low
resolution~\citep[see also][]{Commercon2008}.  However, this problem is resolution dependent, and 
\citet{Governato2004} illustrated this point by presenting disk galaxy
formation simulations in a $\Lambda$CDM context -- without invoking strong feedback -- to show that 
angular momentum loss could be substantially reduced by increasing the mass
and spatial resolution.  Similarly, \citet{Kaufmann2007} used idealized inside-out disk
formation simulations to show that while spurious hydrodynamical
angular momentum loss dominates at low particle resolutions, using $> 10^6$ SPH
particles in each simulated galaxy can make the unphysical
hydrodynamical torques subdominant.

Unfortunately, the very high resolution criteria specified in~\citet{Kaufmann2007} 
make the near-term feasibility of carrying out full cosmological box simulations 
with standard SPH poor.  As a result, many recent galaxy formation studies have adopted 
the ``zoom-in'' technique, where a single galaxy can be simulated at a very 
high mass and spatial resolution~\citep[e.g.][]{Guedes2011, Agertz2011}.  Some 
of these efforts have led to the formation of galaxies that share 
many properties in common with our own disk-dominated Milky Way.   
However,~\citet{Agertz2011} 
argued that -- even though they were successful in reproducing a Milky Way 
type disk galaxy -- the properties of their simulated galaxies depend heavily 
on the choice of the star formation threshold, formation efficiently, feedback 
parameters, and other poorly constrained star formation related parameters.
Since ``zoom-in'' simulations are limited in their scope to one halo at a time, 
it becomes difficult to judge if the same 
simulation parameters (e.g., star formation threshold, etc.) would validly 
reproduce the wide range of observed galaxy morphologies or 
observationally constrained quantities such as the 
global star formation rate at different redshifts.  
So, while ``zoom-in'' simulations are a very useful numerical tool to understand 
how individual galaxies form and evolve, it is necessary to perform large cosmological 
box simulations, where a wide variety of structures should naturally form which 
can then be compared directly to the wealth of observational galaxy data.
Simulations of representative samples of the 
Universe permit us to test more clearly the impact of poorly constrained simulation
parameters on structure formation by exploring the evolution of a full ensemble of 
galaxies, rather than one individual object.  Once it becomes feasible to produce 
a large ensemble of realistic galaxies in a cosmological context, we will have a powerful tool 
to address questions about the driving forces behind galaxy morphological 
evolution that would complement the efforts of ``zoom-in'' simulations.

Another issue that exists in standard density SPH is the  
formation of dense gas ``blobs"~\citep{Kaufmann2006, vandeVoort2011,
  Keres2009letter, vandeVoort2012}, that form via numerical thermal instability 
that occurs in the absence of thermal conductivity~\citep{Hobbs2012}. For 
example,~\citet{Kaufmann2006} presented simulations of inside-out disk 
formation and found a population of dense gas blobs efficiently accreted onto their 
central forming galaxy. These blobs -- which are not found in 
adaptive mesh refinement simulations~\citep{Joung2012} or more modern
SPH algorithms where entropy mixing is included via thermal conductivity~\citep{Hobbs2012}
-- deliver a substantial 
amount of gas to forming galaxies, making them capable of impacting 
the structural properties of galactic gas disks~\citep{Sijacki2011, Hobbs2012}.

One way to improve the prospects of carrying out reliable cosmological
simulations without relying on substantial increases in available computational power is to improve the accuracy of the hydro solver for a fixed resolution or computational cost.  For example, the primary reason for the
required high resolution in~\citet{Kaufmann2007} is to decrease the importance of spurious
hydrodynamical torques that occur at sharp density boundaries between dense
galactic gas disk and the surrounding hot gas haloes.  However, if one could remove
the source of the spurious hydrodynamical torques and improve the thermal 
mixing properties, then it may be possible to 
relax the high resolution criteria to a more attainable level.   It is
possible that this could be achieved by either modifying the SPH
algorithm~\citep[e.g.][]{RitchieThomas2001,PriceThermalConductivity,WadsleyMixing2008,
VSPH,SPHS,rpSPH,SaitohMakino2012,Hopkins2012} or by moving to a grid based code where
these spurious hydrodynamical torques are not expected to
occur~\citep[e.g.][]{Okamoto2003}.   Indeed, in a recent study by
  ~\citet{Aquila} it has been shown for a single galaxy simulated at a
  high resolution via a zoom-in technique that the choice for the adopted 
  hydro solver can impact the galaxy morphology.

In this paper, we explore the formation of gas disks in two cosmological
simulations:  one using a traditional density based SPH formulation as
implemented in {\small GADGET} \citep{gadget2} and one using a novel
moving-mesh grid-based hydrodynamical solver as implemented in {\small
AREPO}~\citep{arepo}.  Both {\small GADGET} and {\small AREPO} are massively
parallel hydrodynamical simulation codes that use the same gravity solver and
sub-grid physics, allowing us to isolate the impact of the hydro solver at an
equivalent number of initial resolution elements and nearly equivalent computational
cost.  We study cosmological simulation runs with the two codes and find
that the gas disk scale lengths associated with the cold gas disks formed in {\small AREPO} are
systematically larger than their {\small GADGET} counterparts.  We discuss the
reasons for these differences including the impact of the numerical resolution. The work
described here is an extension of the analysis presented in \citet{Keres2011}
which discussed the properties of galaxies and haloes.  In this work, we focus
specifically on the properties of gas disks that form within the two
simulations and compare their structural properties.

This paper is structured as follows.  In Section 2, we summarize the 
numerical methodology.  We then present our results for the gas disk 
properties in Section 3, and discuss the causes for the differences in 
Section 4.  Finally, we summarize our findings in Section 5.



\begin{figure*}\centerline{\hbox{
\includegraphics[width=7.2in]{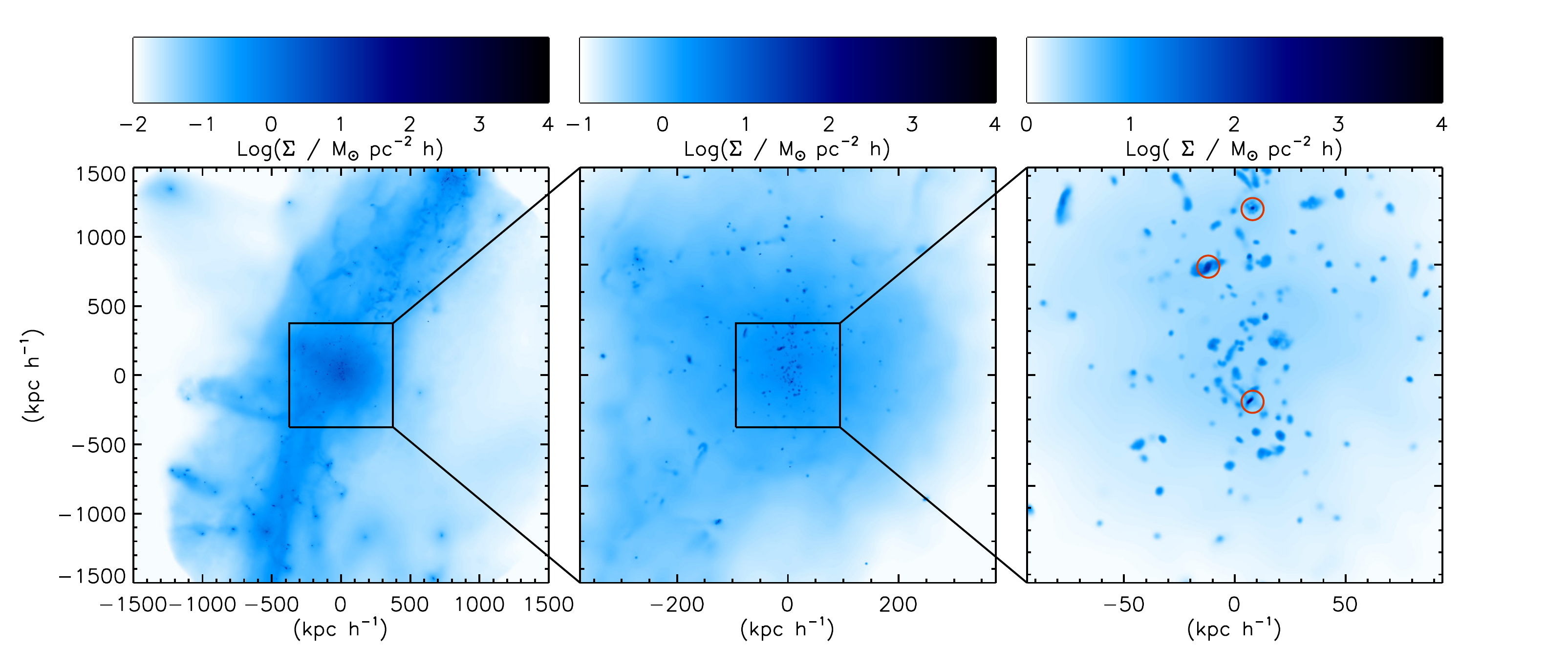}}}
\caption{Maps of the projected gas surface density for one 
object in the {\small GADGET} simulation at redshift $z=1$.  The central 
object has a halo mass of $M=2\times10^{12} h^{-1} M_\odot$.  Three nested views are shown 
to give a clear picture of the gas distribution over a large range of spatial 
scales.  In the rightmost panel, the gas distribution around the central 
galaxy can be seen to be fairly clumpy and the galaxies themselves appear 
fairly compact.  In this image, three galaxies are 
in the process of merging, which we have identified with red circles.  It is helpful to
 directly compare this plot to Figure~\ref{fig:ArepoCenterpiece}, which shows 
 the same maps for the {\small AREPO} simulation.}
\label{fig:GadgetCenterpiece}
\end{figure*}

\begin{figure*}\centerline{\hbox{
\includegraphics[width=7.2in]{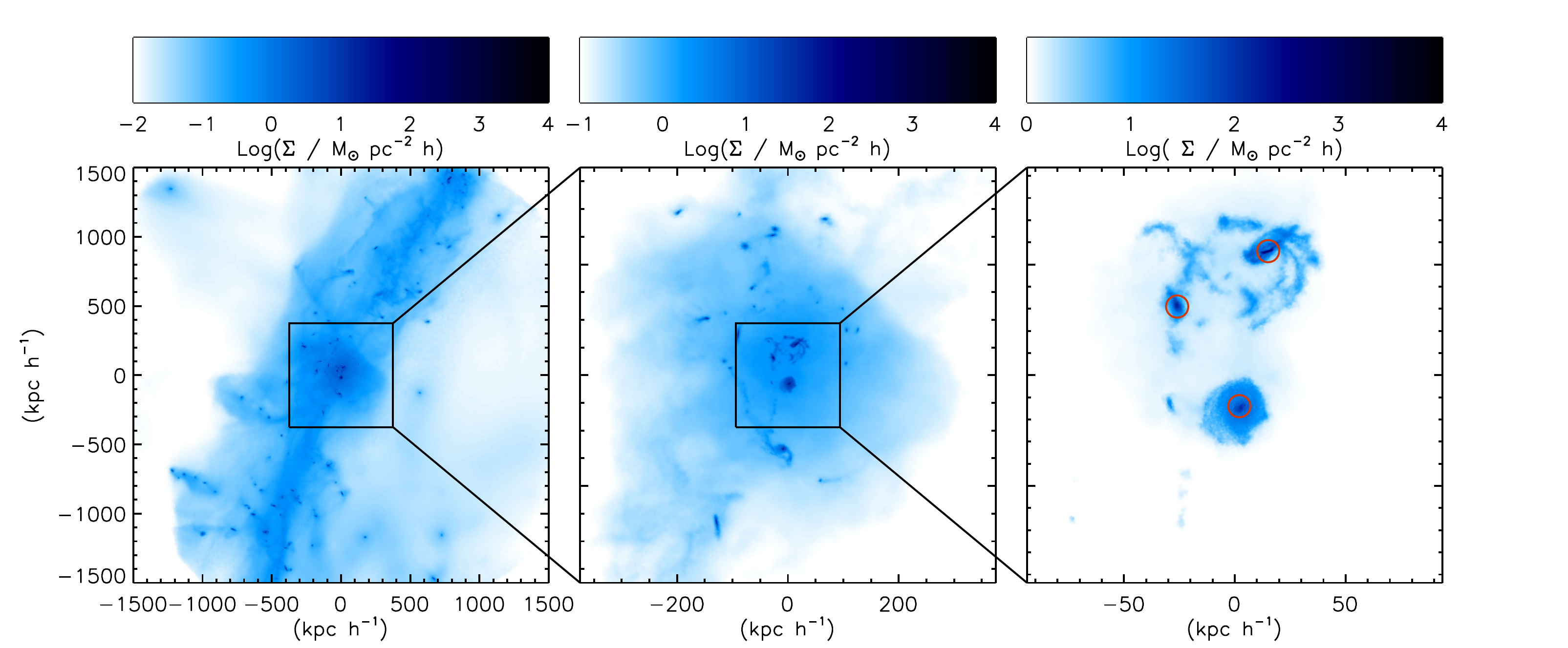}}}
\caption{Same as Figure~\ref{fig:GadgetCenterpiece}, but for the 
{\small AREPO} simulation.  In contrast to Figure~\ref{fig:GadgetCenterpiece}, the 
gas distribution around the central galaxy in the rightmost panel has much fewer gas 
clumps.  The three galaxies that are in the process of merging are highlighted with red circles.}
\label{fig:ArepoCenterpiece}
\end{figure*}

\begin{figure*}\centerline{\hbox{
\includegraphics[width=7.2in]{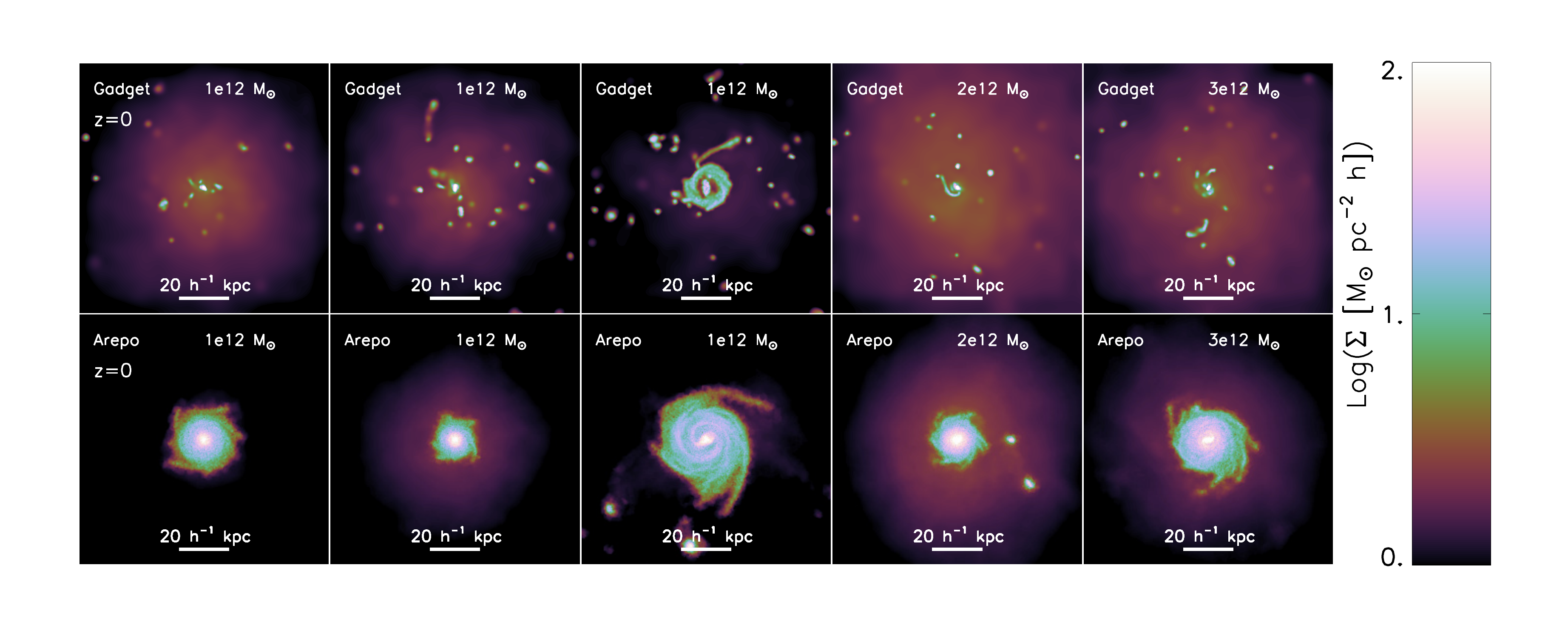}}}
\caption{Projected gas surface density maps of $5$ matched objects in {\small
    GADGET} and {\small AREPO} chosen  at redshift $z=0$ with host halo masses
  $\sim 10^{12} h^{-1} {\rm M}_\odot$. There are clear differences in the  extent of
  the central gas disk.  In addition, the prevalence of dense gas blobs is
  much higher in the {\small GADGET}  simulation.}
\label{fig:PostageStamps_z0_blobs}
\end{figure*}

\begin{figure*}\centerline{\hbox{
\includegraphics[width=7.2in]{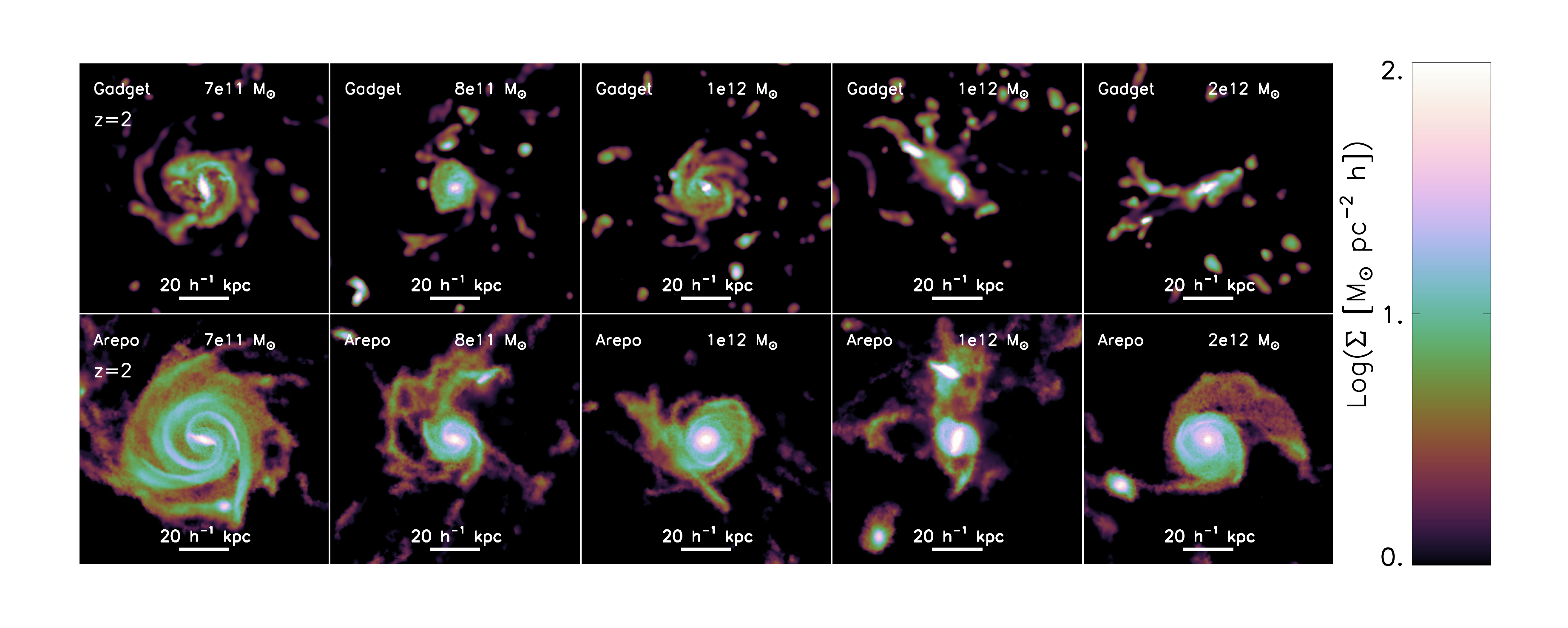}}}
\caption{Projected gas surface density maps of $5$ matched objects in {\small
    GADGET} and {\small AREPO} chosen  at redshift $z=2$ with host halo masses
  $\sim 10^{12} h^{-1} {\rm M}_\odot$.  There are clear differences in the  extent of
  the central gas disk.  In addition, the prevalence of dense gas blobs is
  much higher in the {\small GADGET} simulation.}
\label{fig:PostageStamps_z2_blobs}
\end{figure*}

\begin{figure*}\centerline{\hbox{
\includegraphics[width=16.9truecm]{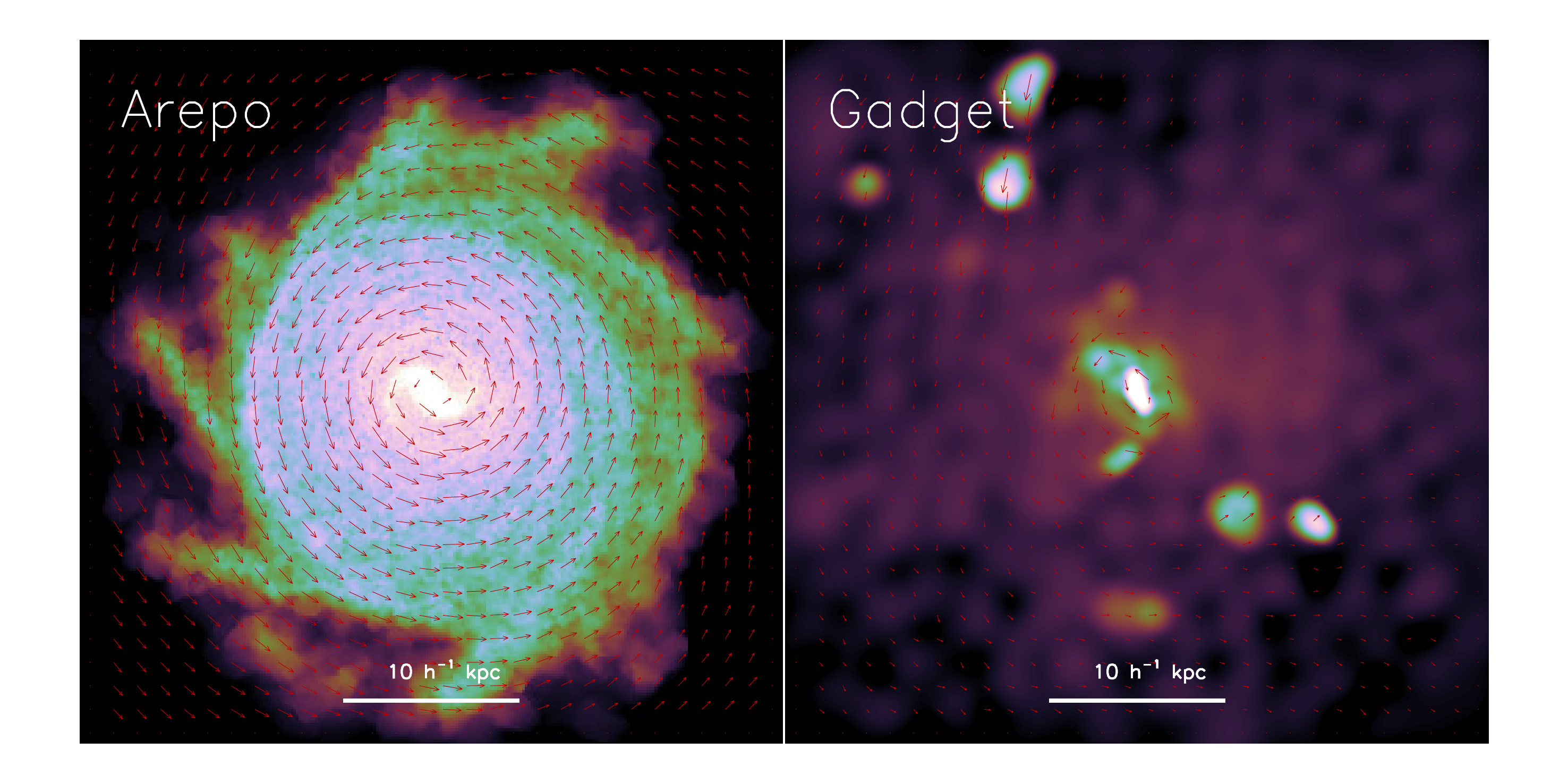}}}
\caption{Maps of the projected gas surface density for a typical matched
  galaxy in a $M_{ {\rm Halo} } = 10^{12} M_\odot$ halo. Red overplotted arrows denote the
  local gas velocity field. The AREPO galaxy (left panel) is significantly
  more rotationally supported than its GADGET counterpart (right panel).}
\label{fig:velocity_maps}
\end{figure*}

\begin{figure*}\centerline{\vbox{\hbox{
\includegraphics[width=5.9truecm,height=5.9truecm]{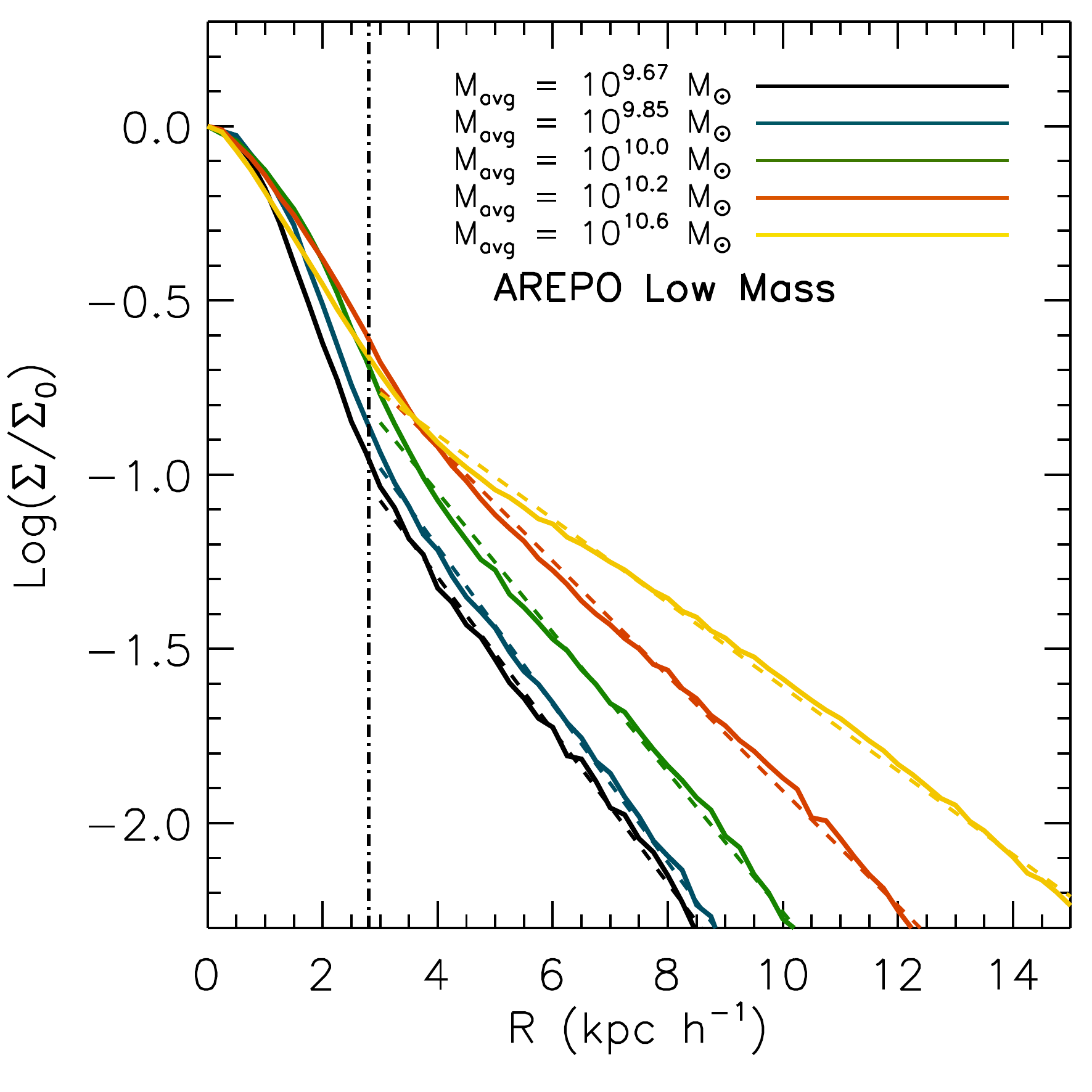}
\includegraphics[width=5.9truecm,height=5.9truecm]{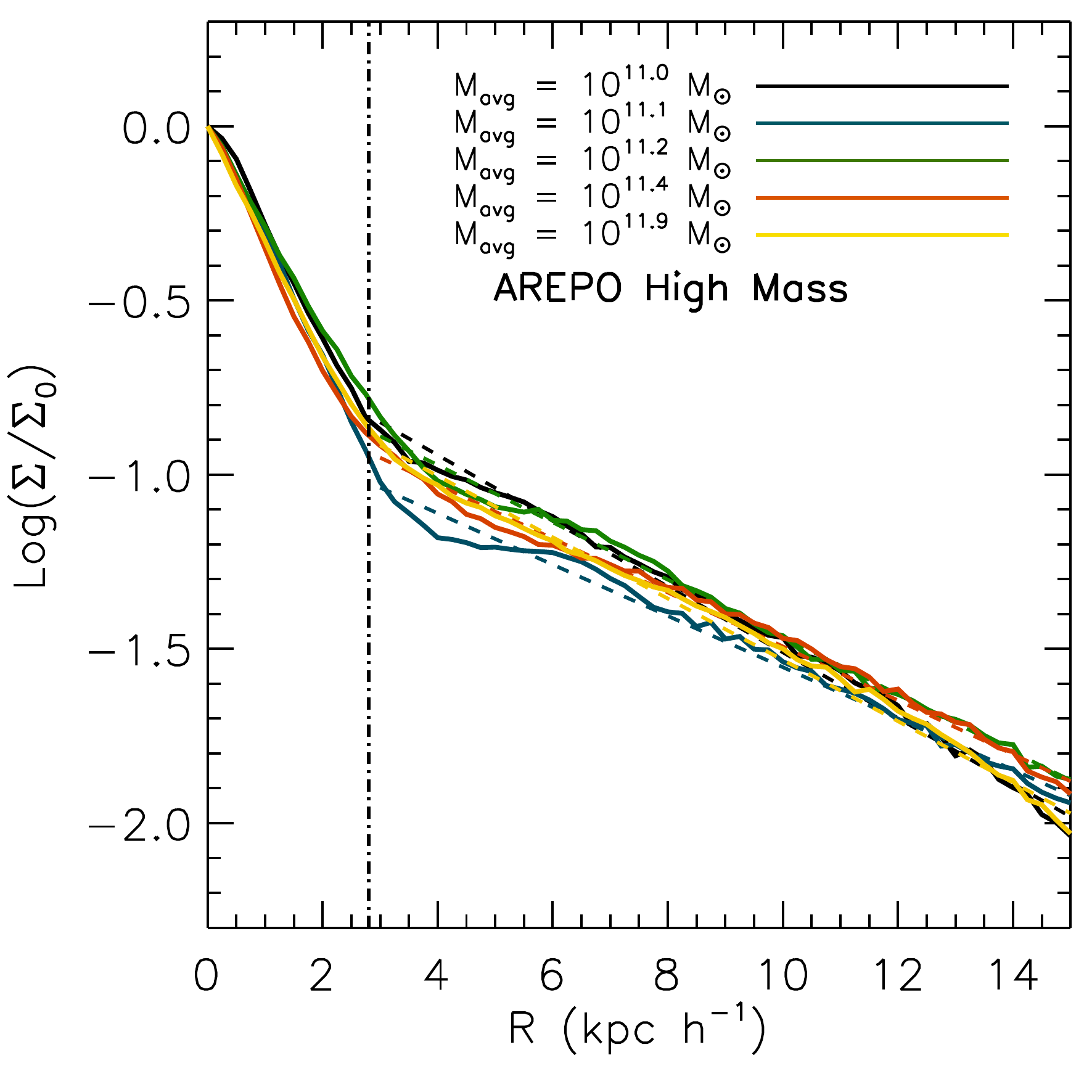}}
\hbox{
\includegraphics[width=5.9truecm,height=5.9truecm]{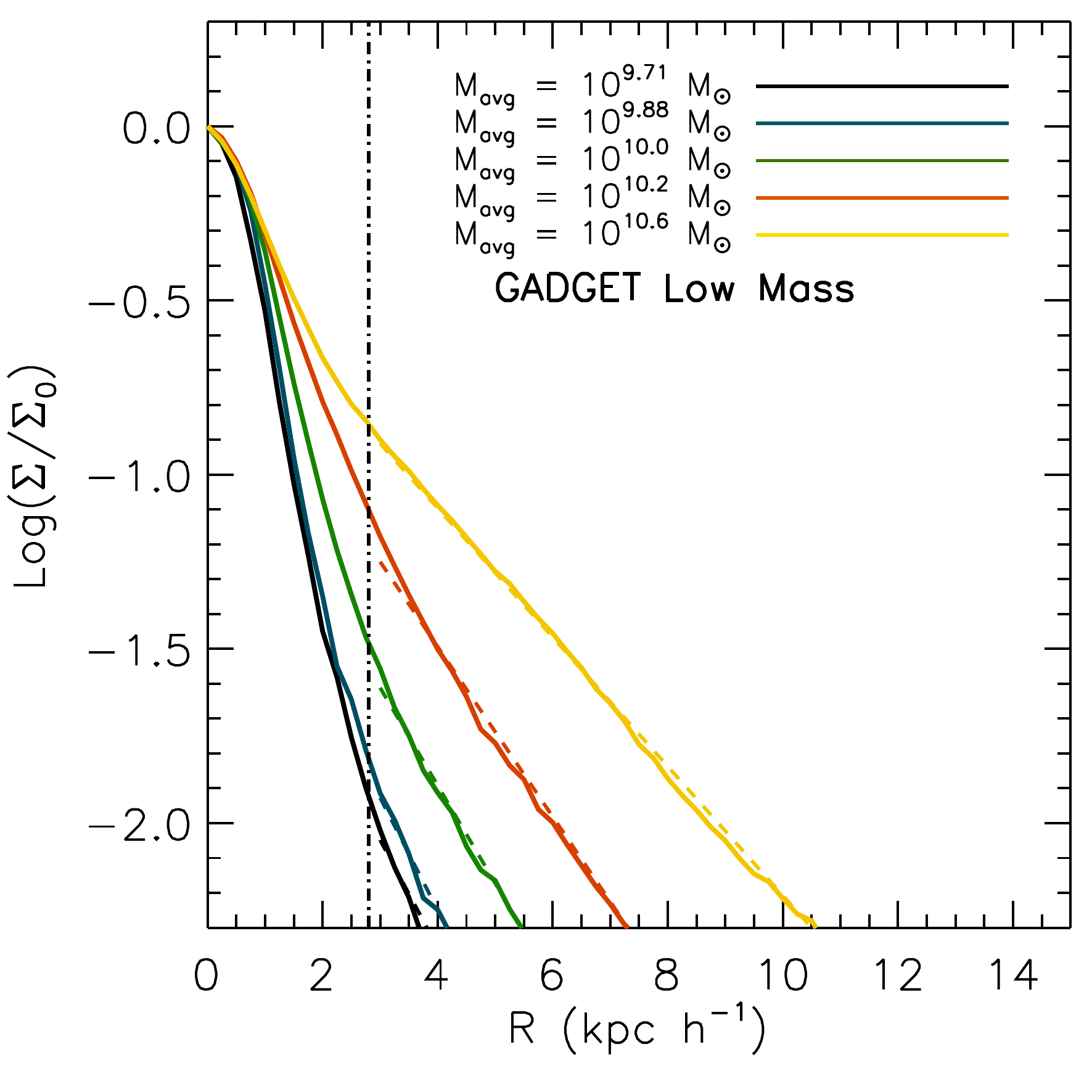}
\includegraphics[width=5.9truecm,height=5.9truecm]{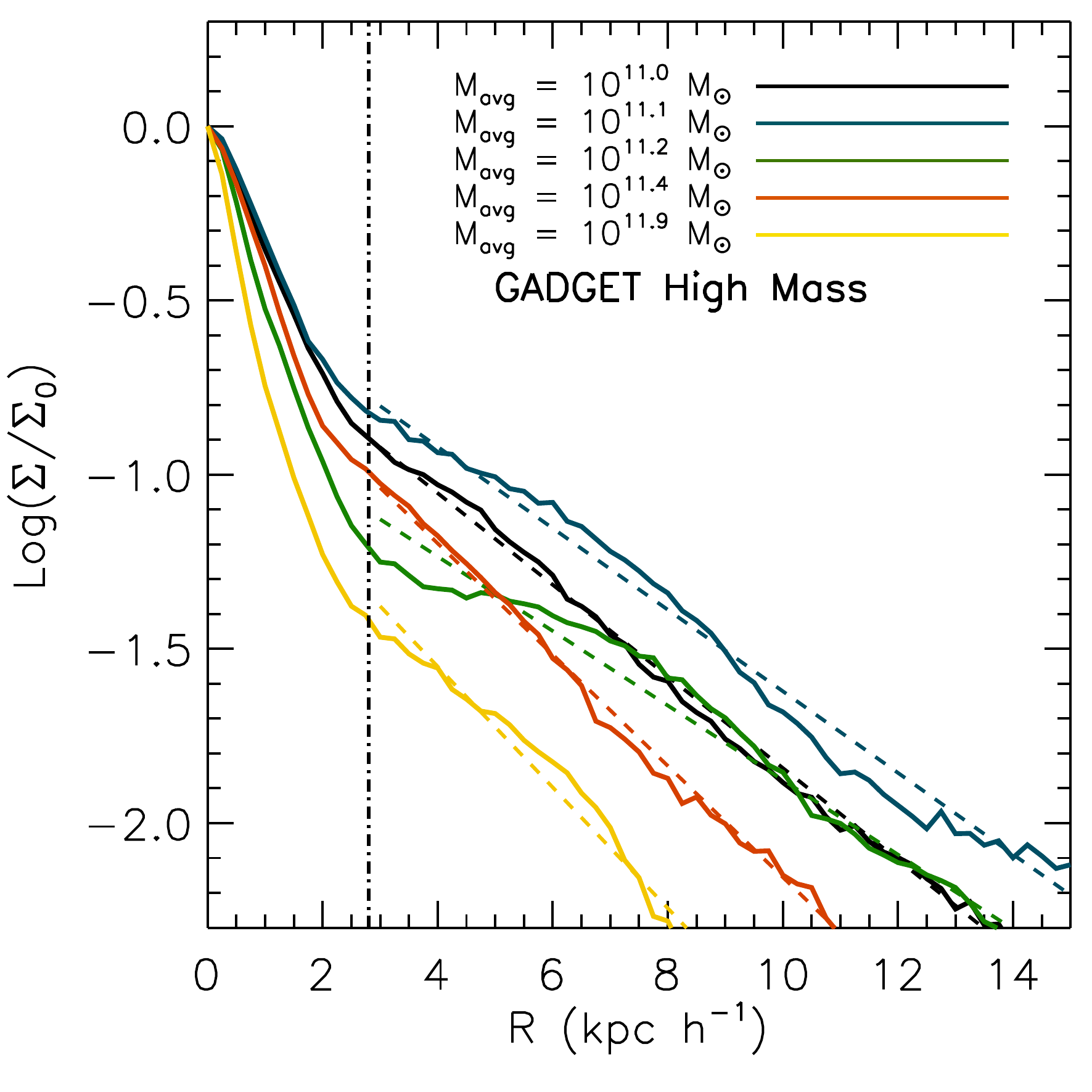}}}}
\caption{Stacked mass profiles at $z=0$ for the low mass {\small AREPO}
  objects (top left), high mass {\small AREPO} objects (top right), low mass
  {\small GADGET} objects (bottom left) and high mass {\small GADGET} objects
  (bottom right). The true mass profiles are shown as solid lines, while the
  best fit exponential functions are shown as dashed lines. Each colour
  corresponds to an average surface density profile for all galaxies in a
  given halo mass range, with the average halo mass, $M_{\rm avg}$, indicated
  in the legend. The vertical dot-dashed lines denote the spatial scale over
  which the gravitational potential is softened, i.e. $2.8\, h^{-1} \,{\rm
    kpc}$, co-moving.}
\label{fig:TrueSDProfiles}
\end{figure*}

\section{Methods}

We consider the simulations presented in \citet{Vogelsberger2011}, which were 
performed with the cosmological codes {\small GADGET}~\citep{gadget2} and
{\small AREPO}~\citep{arepo}.  These simulations follow structure formation in
an $L = 20 \,\mathrm{Mpc} \; h^{-1}$ box assuming a standard $\Lambda$CDM
cosmology ($\Omega_\Lambda = 0.73$, $\Omega_0 = 0.27$, $h=0.7$,
$\sigma_8=0.8$). The simulations contain $N_{ {\rm DM} } = 512^3$ dark matter
particles and an equivalent number of gas resolution elements at the start of
the simulation, giving a mass resolution of $m_{ {\rm DM} } = 3.722 \times
10^6 h^{-1} M_\odot$ and $m_{{\rm gas}} = 7.444 \times 10^5  h^{-1} M_\odot$
(note that the gas resolution element mass can change in {\small AREPO} due
to mass advection across cell boundaries, but all cells maintain a mass
within a factor of 2 of this target value). A comoving gravitational
softening length of $\epsilon = 1 h^{-1} \; {\rm kpc} $ was used.

\subsection{Comparison Parameter Selection}

A central premise for our comparison is that {\small GADGET} and {\small 
AREPO} are very similar simulation codes which allows us to hold 
a large number of simulation parameters and procedures the same.  In addition
to containing the same gravity solver, the prescriptions for the radiative
cooling of gas~\citep{KatzCooling},
the evolution of the ionizing background radiation
field~\citep{FG2009}, and star formation with associated
feedback~\citep{SHMultiPhase} are identical.  The fundamental difference
between the codes lies in their handling of gas hydrodynamics.  While {\small
GADGET} uses a standard density based smoothed particle hydrodynamics (SPH)
technique to evolve the gas, {\small AREPO} solves the fluid equations using
an exact Riemann solver on a moving unstructured mesh based on a Voronoi
tessellation.
{\small AREPO} has a number of advantages over the standard
SPH method due to its ability to, e.g., capture shocks more accurately and 
better resolve fluid instabilities~\citep{arepo,Sijacki2011,
SpringelSPHBookChapter}.  In particular, {\small AREPO} 
can handle weak shocks more reliably than standard SPH
codes which wash these features out because of the effects of artificial
viscosity~\citep[see, e.g.,][]{Keshet2003}.
Also, with respect to Eulerian adaptive mesh
refinement (AMR) codes used in cosmology, {\small AREPO} employs a more
accurate gravity solver~\citep[see, e.g.,][]{OShea2005}
and a spatial refinement that is continuous with the
motion of the fluid. 
Although these traits have been demonstrated in isolated
test problems~\citep{arepo,Sijacki2011}, we would like to understand how
these numerical effects can impact the evolution of galaxies in cosmological
simulations.  Furthermore, because the same gravity solver and sub-grid
physics prescriptions are used, we avoid some of the uncertainties that remain
in code comparisons where these are allowed to
vary~\citep[e.g.,][]{FrenkSBC1999}.

For our comparison, we choose to have both codes start from identical initial
conditions and have the same initial number of resolution elements ($N_{\rm 
DM}=512^3$, $N_{\rm gas}=512^3$).  From a practical standpoint, using
the same number of initial resolution elements leads to comparable computational 
cost for the two codes (with the {\small AREPO} simulation taking $\sim$ 30\% 
more CPU time).  This is an important consideration, because it is the CPU 
expense which sets limits on the size and complexity of simulations that can 
be run.  However, since the two codes have roughly comparable CPU 
consumption for the same number of resolution elements, we simply 
note that neither code has a distinct advantage in this area.

From a more physical standpoint, using the same number of 
initial resolution elements results 
in similar mass resolutions between the two codes.  For the dark matter 
component where no inter-particle mass transfer is required, both 
simulations share identical mass resolutions.  For the baryon component, 
the SPH particles in {\small GADGET} have a fixed mass in time while 
the cells in {\small AREPO} have a time dependent mass due to mass 
advection to their neighbors when solving the Riemann problem
across cell boundaries.  We note, however, that we have included 
a refinement/de-refinement 
scheme that maintains all hydro cells in our {\small AREPO} simulation 
within a factor of two of the SPH particle mass in our {\small GADGET} 
simulation~\citep{Vogelsberger2011}.  As a result, both simulations 
share similar (although not identical) mass resolutions.  As
discussed in~\cite{Vogelsberger2011}, we emphasize
that the fixed particle mass in SPH is connected to
inaccuracies made in solving the continuity equation, preventing
this algorithm from correctly handling mixing.


Since the dark matter components in both simulations 
evolve similarly (with some potential differences resulting from the 
influences of the 
baryon components), the same large scale structure and halo 
properties are present in both simulations~\citep{Vogelsberger2011}.  
As a result, we can compare gas disks which reside in a set of matching 
haloes that are identified as being present in both simulations.  
Figures~\ref{fig:GadgetCenterpiece} and~\ref{fig:ArepoCenterpiece} illustrate 
this point for two matching haloes taken from our simulations.  In these 
nested maps of the projected gas surface density, it can be seen that the 
distribution of gas on large scales is similar in the two codes (see
the leftmost panels of Figures~\ref{fig:GadgetCenterpiece} 
and~\ref{fig:ArepoCenterpiece}).  This is an expected result, as 
the gas distribution traces the dark matter distribution on large scales.  
However, as one examines the distribution 
of gas on galactic scales (as shown in the central and rightmost panels), 
it becomes clear that there are prominent differences between the two codes.  
Specifically, while the gas in the {\small GADGET} simulation is distributed in a large number 
of compact and dense clumps, in {\small AREPO} gas has a much smoother distribution.
Further inspection of 
Figures~\ref{fig:GadgetCenterpiece} and~\ref{fig:ArepoCenterpiece} 
shows that in the central region there are three galaxies -- highlighted by 
red circles -- in the process of merging.  The effects of their 
mutual interaction can be seen in the case of {\small AREPO}, where 
tidal features are visible.  Although these galaxies are present in {\small GADGET}
as well, tidal features are much less prominent because the gas is more 
centrally concentrated and less rotationally supported. Perhaps the most
striking finding from Figures~\ref{fig:GadgetCenterpiece} 
and~\ref{fig:ArepoCenterpiece}, results from a comparison 
of the galaxy located nearest to the origin in these plots.  While this object
appears as a smooth, spatially extended disk in {\small AREPO}, the 
same object in {\small GADGET} is better described as a featureless blob.

Figures~\ref{fig:GadgetCenterpiece} and~\ref{fig:ArepoCenterpiece} summarize 
the motivation for this work.  Even though these simulations 
have been initiated from the same initial conditions, share the same 
feedback prescriptions, and use the same 
number of initial resolution elements, the detailed morphological properties 
of the gas distribution on galactic scales can be very different.  This is a 
very important point, because it indicates that the hydro solver has a significant 
impact on the gas properties.  While 
these differences can be fairly easily identified from the gas surface 
density maps, a primary goal of this paper is to produce a detailed quantitative comparison 
of the sizes of the gas disks 
based on a large sample of galaxies matched between the two simulations.

\subsection{Gas Disk Analysis}

In what follows, we contrast gas disks that form in the {\small GADGET}
and {\small AREPO} cosmological simulations.  To facilitate this comparison, 
we first identify a sample of matched galaxies from the two simulations.  
We start by building a catalog of all structure in the each simulation independently 
using the {\small SUBFIND} tool~\citep{subfind}.  We assemble a population of 
``matching'' haloes by finding objects in the {\small GADGET} and {\small AREPO} 
simulations that have the same total mass to within 10\% and potential minimum 
locations that are not offset by more than 25\% of their half mass radii. 
We remove any pairs where the centre of mass is offset by more than $10\,h^{-1} 
{\rm kpc}$ from the most tightly bound particle in either halo, as this may be an 
indication of a merging system.  These selection criteria yield $1367$ matching 
haloes in both simulations with total halo masses above $10^{10}  h^{-1} \, 
{\rm M_\odot} $ at redshift $z=0$.  Figures~\ref{fig:PostageStamps_z0_blobs} 
and~\ref{fig:PostageStamps_z2_blobs} show examples of the local gas 
density around a series of five such matching objects 
in {\small GADGET} and {\small AREPO} at redshift $z=0$ and $z=2$, respectively.

In addition to showing differences in the spatial distribution of gas,
  the gas  component of galaxies formed in {\small GADGET} and {\small AREPO}
  also differ kinematically.  Figure~\ref{fig:velocity_maps} shows the maps of
  the gas surface density  for an example matched galaxy with vectors
  indicating the local gas velocity field.  Based on the topology of the
  velocity field, it is evident that the large gas disk in {\small AREPO} shows
  clear rotation about the disk's center. In contrast, the velocity field of
  the central galaxy for the same halo in the {\small GADGET} simulation, exhibits
  much lower level of circulation.

To analyze the gas disk properties, we identify the gravitationally 
bound cold and dense 
gas within each halo.  We distinguish the diffuse hot halo from the colder, more dense 
rotating gas disk by making a cut in the $T-\rho$ phase diagram at
\begin{equation}
\log_{10}\left(\frac{T} { \mathrm{[K]}}\right) =  6 + 0.25 \log_{10}\left(\frac{ \rho}{ \; \mathrm{10^{10} [M_\odot \;} h^2 \mathrm{\; kpc^{-3}]} } \right) .
\label{eqn:GasCut}
\end{equation}
Our subsequent analysis depends on first removing the hot halo
component before measuring the disk surface density profile.
However, our results are not very sensitive to this
particular cut in the phase diagram (i.e. moving the normalization of
this cut up or down by a factor of 5 would not change our
conclusions). The remaining cold and dense gas is translated to place
the centre of mass at the origin and rotated to align the net gas
angular momentum vector in the $\hat z$ direction.

Before moving forward, we note that the definition of cold/dense
gas cut defined in equation~(\ref{eqn:GasCut}) can select material which
is not part of the central gas disk.  In particular, we find that high mass
haloes in {\small GADGET} contain a population of 
low mass cold gas clumps which have no associated dark
matter overdensity.  The rightmost panel in Figure~\ref{fig:GadgetCenterpiece}
shows an example of these gas clumps outside of the central object and
a similar population of clumps can be found in
all massive (i.e. $\sim 10^{12} h^{-1} M_\odot $) haloes in the {\small
GADGET} run.  We note that such clumps are also seen in 
simulations of galaxy formation carried out with other SPH 
codes~\citep[e.g.][]{Okamoto2008, Guedes2011}.  Most of these clumps are not part of the galactic
disk -- many of them have entered the halo for the first time,
have not yet had any contact with the central galaxy, and are on
very non-circular trajectories.  However, the density and temperature of these
clumps allows them to be selected as ``disk mass'' according to the definition
outlined in equation~(\ref{eqn:GasCut}).  

In principle, we could remove these cold gas clumps by imposing some
additional criteria in our ``disk gas'' selection, e.g., we could require disk
gas to be on nearly circular trajectories or link together the central disk
using a friends-of-friends (FOF) algorithm.  Nonetheless, at this point we
choose not to impose any additional selection criteria for two reasons.
First, we find that de-selecting clump material can be sensitive to the
details of the clump removal technique that we use.  For example, while running
a FOF algorithm on the cold and dense gas can efficiently remove clumps that
are more than 15 or 20 $h^{-1}$ kpc from the central galaxy, removing the
most centrally located gas clumps can depend on our choice for the linking
length.  Second, since these clumps are only present in {\small GADGET}, we
find that removing them tends to decrease the disk scale lengths obtained for
massive {\small GADGET} galaxies without having any noticeable impact on the
same objects in {\small AREPO}.  Since one of the conclusions in this paper is
that gas disks formed in {\small GADGET} are indeed more centrally
concentrated than those formed in {\small AREPO}, we have tried to avoid any
steps in our analysis that could be perceived as artificially pushing us
toward that result.  Thus, in the following section on gas disk properties, we
leave these clumps in our definition of disk mass and save additional
discussion about their origin and impact for section~\ref{sec:ColdClumps}.

\section{Disk Comparison}
\label{sec:Results}
\label{sec:ExpFits}

To find the average surface density profile for gas disks, we stack the 
profiles for all objects in a given halo mass range. A representative set of
stacked surface density profiles is shown in
Figure~\ref{fig:TrueSDProfiles}. We find the best-fit exponential profile 
\begin{equation}
\Sigma (r) = \Sigma_0 \; \mathrm{exp} (-r/R_{ {\rm d} })
\label{eqn:Cut}
\end{equation}
for each surface density profile via a chi-squared minimization.  Note from
Figure~\ref{fig:TrueSDProfiles} that most of the profiles can be fit using a
single exponential.  However, almost all profiles are distinctly steeper in the
inner regions, and this is especially pronounced for {\small GADGET}
galaxies.  This compact central feature is typically associated with a slowly
rotating spheroidal component.  Since we are primarily interested in the
structure of the spatially extended disks and because this inner region is not
well-resolved in our current simulations (the softening length is $1\,h^{-1}
\, {\rm kpc}$ co-moving, with gravity becoming fully Newtonian after $2.8\, 
h^{-1}\, {\rm kpc}$, co-moving, for the spline softening employed
here \citep{HK89}), 
we perform fits by both including and
excluding the central $2.8\,h^{-1}\, {\rm kpc}$. We check the quality of each
fit by comparing the integral of the best fit surface density profile in the
fitted region to the true disk mass measured in the simulation. We note that
when we exclude the central region all fits for both {\small GADGET} and
{\small AREPO} return the cold, dense gas mass (as defined by
equation~\ref{eqn:Cut}) within 10\%, indicating that our prescribed
exponential functions are serving as good representations of the actual
stacked surface density profiles. When we include the central region, the
{\small GADGET} fits underestimate the disk mass by $\sim$20-30\% while the
{\small AREPO} fits underestimate the disk mass by $\sim$10\%.  This
difference -- which is more pronounced for {\small GADGET} -- 
is caused by a concentration of material in the
central region.

\begin{figure*}\centerline{\hbox{
\includegraphics[width=7.9truecm,height=7.9truecm]{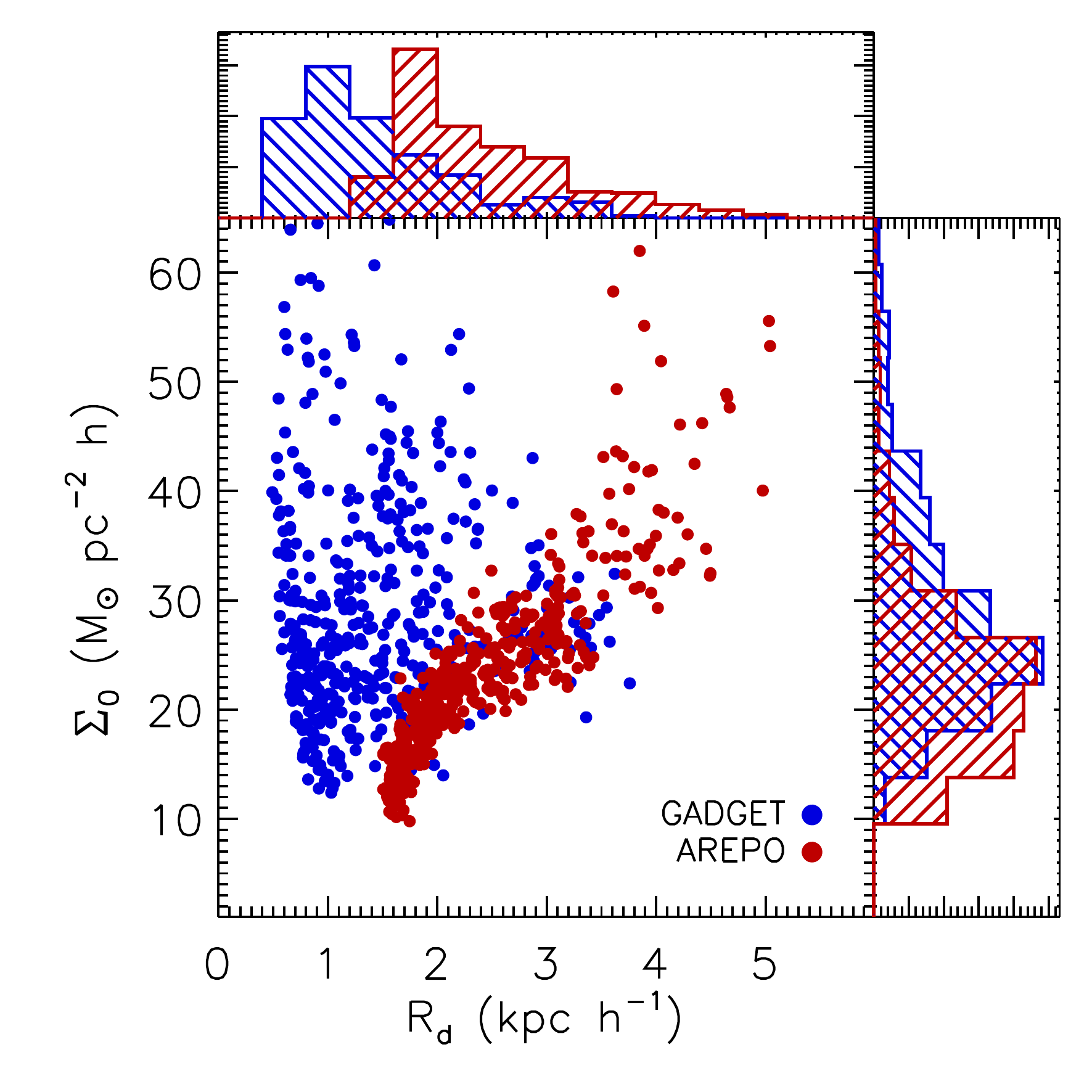}
\includegraphics[width=7.9truecm,height=7.9truecm]{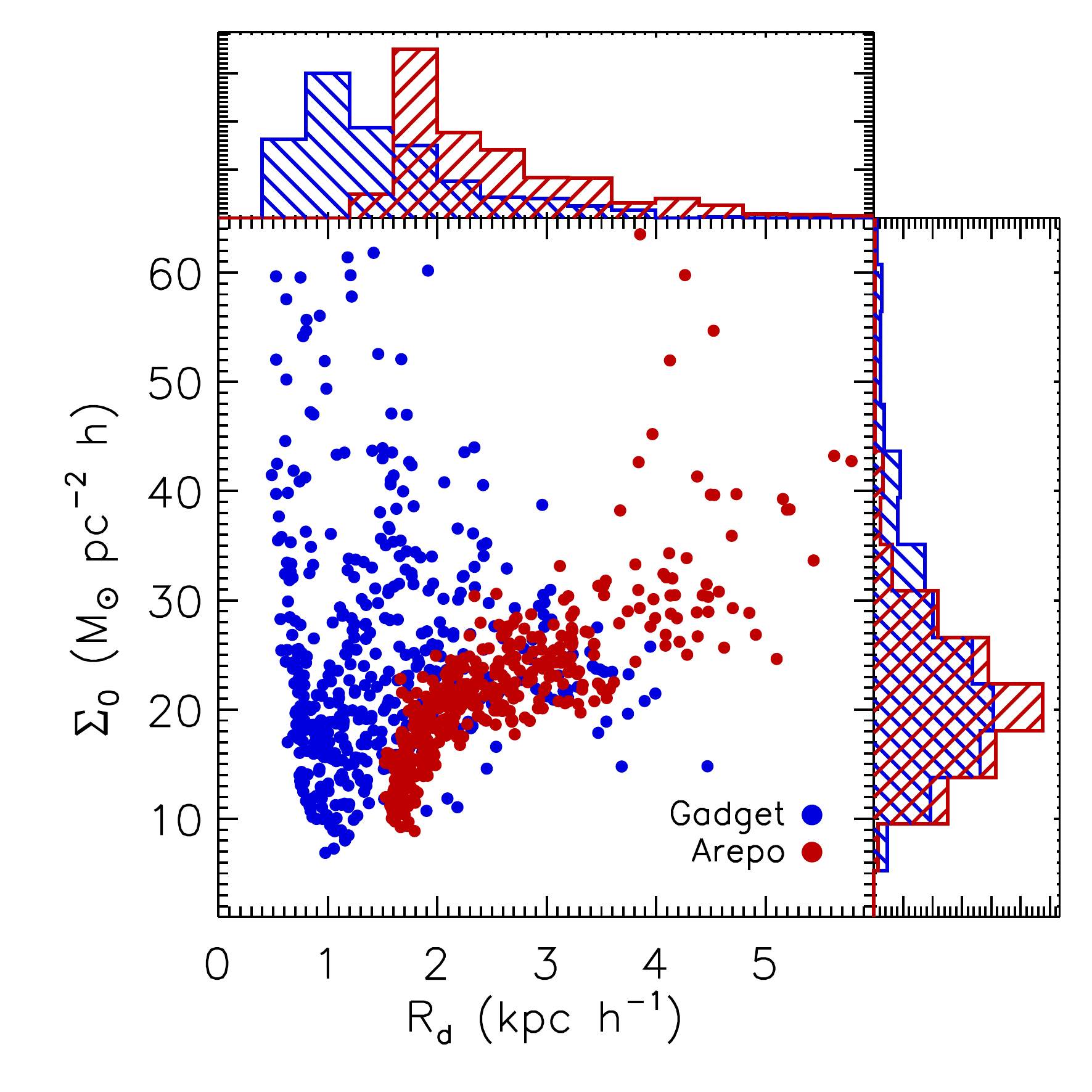}}}
\caption{The best fit exponential surface density profile parameters to the
  whole disk (left) and with the central region excluded (right). The
  distribution of best fit parameters for $R_{\rm d}$ and $\Sigma_{0}$ is
  illustrated in the histograms. Clearly, {\small AREPO} objects have larger
  disk scale lengths than their {\small GADGET} counterparts. Including or
  excluding the central region does not alter this conclusion.}
\label{fig:ExpFits}
\end{figure*}

The distribution of best fit parameters, $\Sigma_0$ and $R_{\rm d}$, of the
exponential surface density profiles are shown in Figure~\ref{fig:ExpFits}.  We
have performed the fits when including (left panel) and excluding (right 
panel) the central region.  We find the {\small GADGET} best-fit central
densities are systematically lowered when we exclude the central region, while 
the {\small AREPO} central densities changes less.   This is a result
of the fact that, on average, {\small GADGET} objects contain a larger
fraction of their gas in the central region.  More importantly, we find that
the distribution of disk scale lengths is systematically different between
the {\small GADGET} and {\small AREPO} galaxies, with the {\small
AREPO} galaxies being described by larger scale lengths regardless of
whether we account for the central region.  This is a central result of this
paper, and confirms the idea that the {\small AREPO} disks are larger
based on visual inspection of the gas surface density maps.  
We emphasize that 
changing the number of galaxies in each stacked surface density
profile, the exact normalization of our cold/hot phase boundary cut, or other
detailed aspects of the analysis does not affect this conclusion.

\begin{figure*}\centerline{\hbox{
\includegraphics[width=7.9truecm,height=7.9truecm]{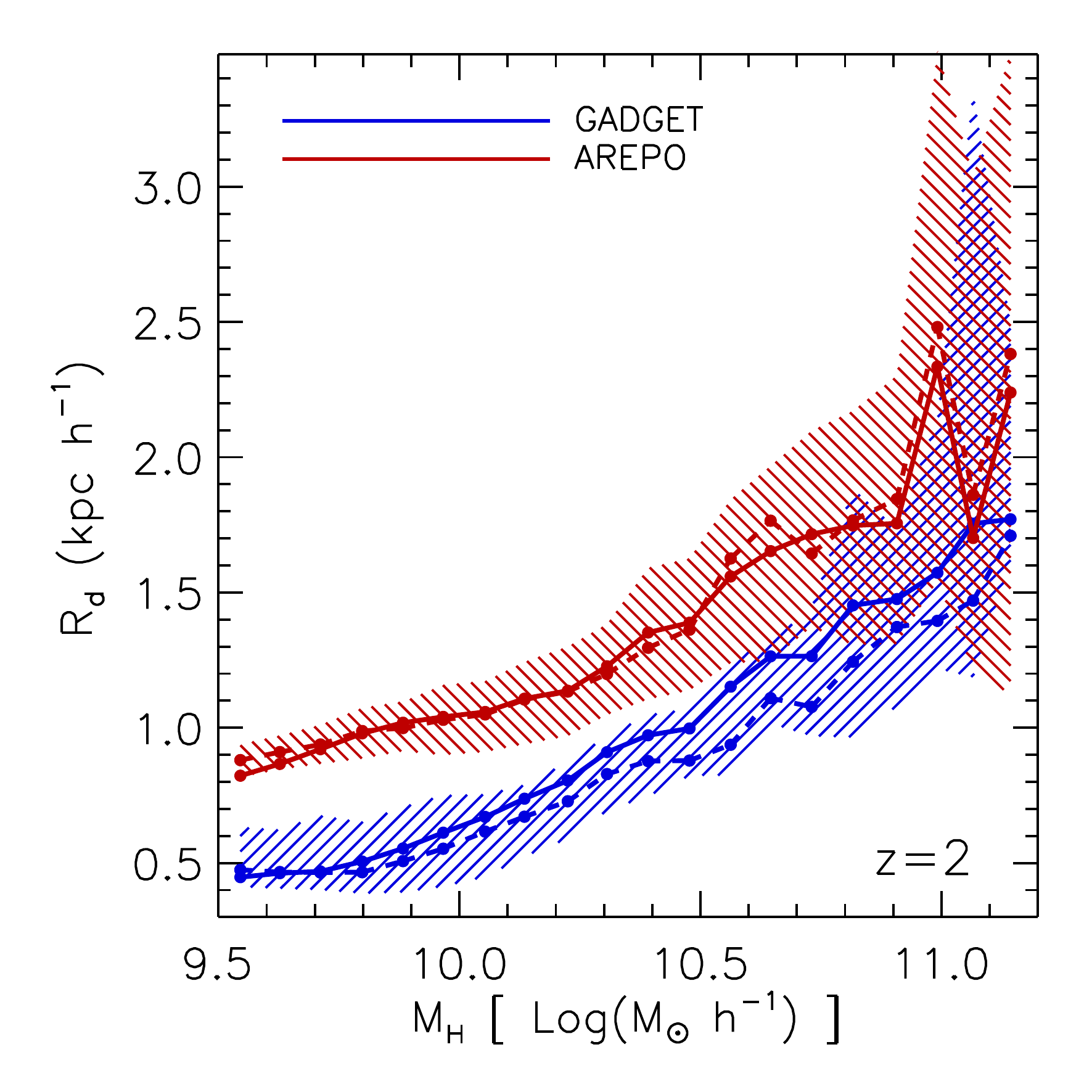}
\includegraphics[width=7.9truecm,height=7.9truecm]{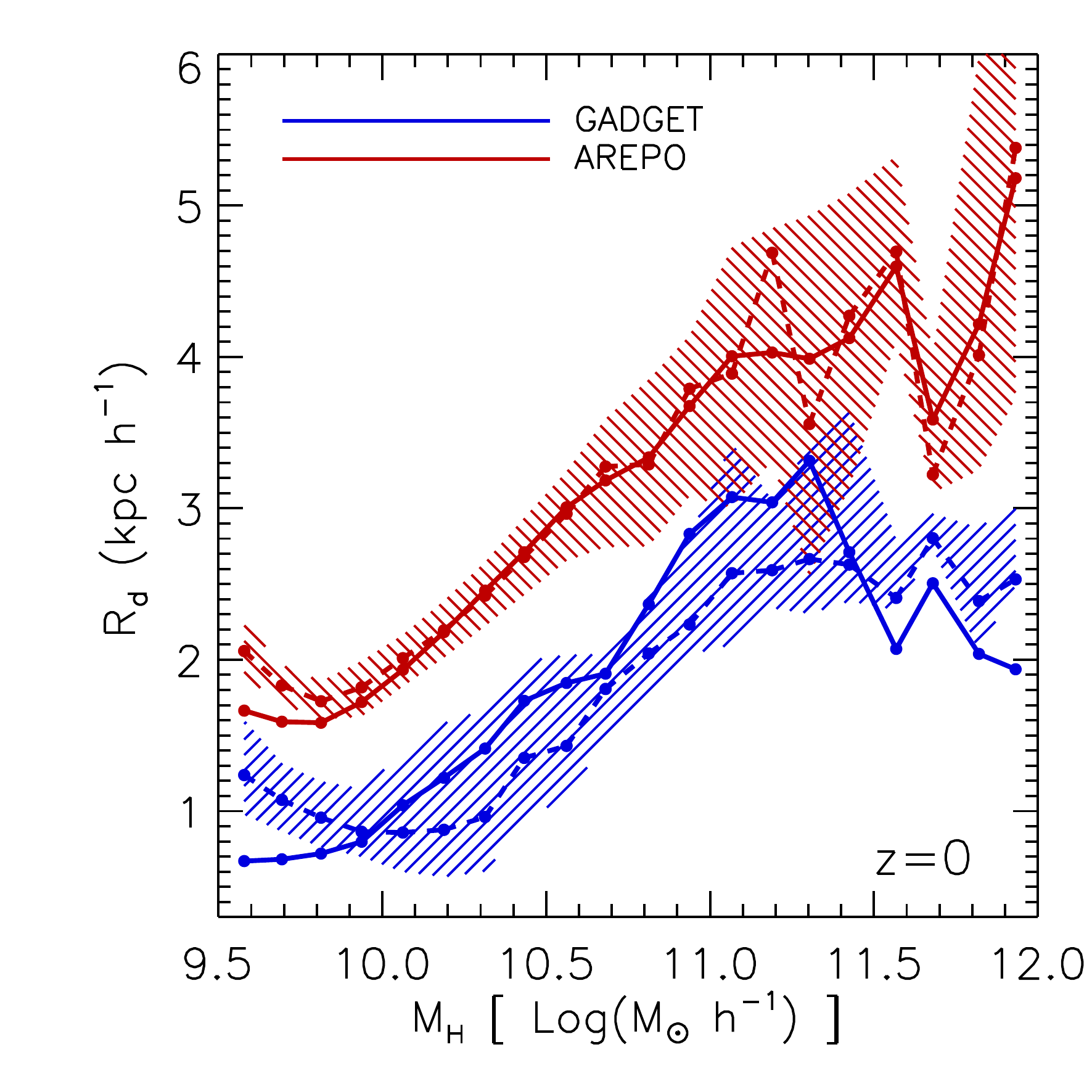}}}
\caption{The best fit exponential disk scale lengths (in physical units) as a
  function of halo mass for {\small GADGET} and {\small AREPO} objects at
  $z=2$ (right) and $z=0$ (left). Continuous lines indicate $R_{\rm d}$ values
  for the stacked galaxies in each mass bin, while dashed lines denote the
  median $R_{\rm d}$ obtained by fitting exponential profiles to individual
  objects. Hatched regions are 25\% and 75\% of the distribution. The {\small
    AREPO} disks are substantially larger at all halo masses and at both
  redshifts.}
\label{fig:ExpFits_z}
\end{figure*}

Figure~\ref{fig:ExpFits} shows clearly the average offset toward larger disk scale lengths for the {\small AREPO} gas disks, we find it instructive to show the best-fit exponential disk scale lengths as a function of host halo mass.  Figure~\ref{fig:ExpFits_z} shows the disk scale lengths (including the central $2.8\, h^{-1}\, {\rm kpc}$) as a function of halo mass at redshifts $z = 2$ and $z = 0$ in the left and right panels, respectively. This allows us to see more clearly how the gas disks from the two codes compare to one another at a fixed halo mass. For each bin we show disk scale length values obtained by first stacking the objects and then fitting the exponential surface
density profiles (continuous lines), and by computing the median disk scale length from a set of individually fit galaxy surface density profiles (dashed lines).  Hatched regions mark 25\% and 75\% of the distribution for the individually fit profiles.  Regardless of the adopted procedure, the {\small AREPO} galaxies have larger scale lengths than their {\small GADGET} counterparts at both redshifts and for all halo masses.  From Figure~\ref{fig:ExpFits_z}, it results that the {\small AREPO} disks are between 1.5 to 2 times larger than their {\small GADGET} counterparts.

We note that for objects with a halo mass $M > 10^{11.5} h^{-1}\; M_\odot $ there
is a discrepancy between the disk scale lengths obtained using our stacking
procedure with respect to the median disk scale lengths obtained from
individual objects in the {\small GADGET} simulations.  The reason for this
discrepancy is the presence of cold gas blobs surrounding the central gas
disk.  As discussed in the previous section, since these blobs are cold and
dense they are included in our definition of ``disk gas''.  Stacking many
objects with a large population of blobs can then systematically bias the
estimate of the disk scale length.  Thus we note that the apparent trend  of
the disk scale length with the halo mass at the high mass end in the {\small
GADGET} simulations as seen in the lower right panel of
Figure~\ref{fig:ExpFits_z} is affected by the presence of cold blobs.  For
these high mass systems, the central disks are just not well described by
single exponential profiles.  An analogous population of cold gas blobs is not
present in the {\small AREPO} simulation.  This is the primary reasons why the
solid and dashed lines -- representing the stacked best fit disk scale length
and median of individually fit disk scale lengths -- are in better mutual
agreement for {\small AREPO} galaxies.

\begin{figure}\centerline{\hbox{
\includegraphics[width=7.9truecm]{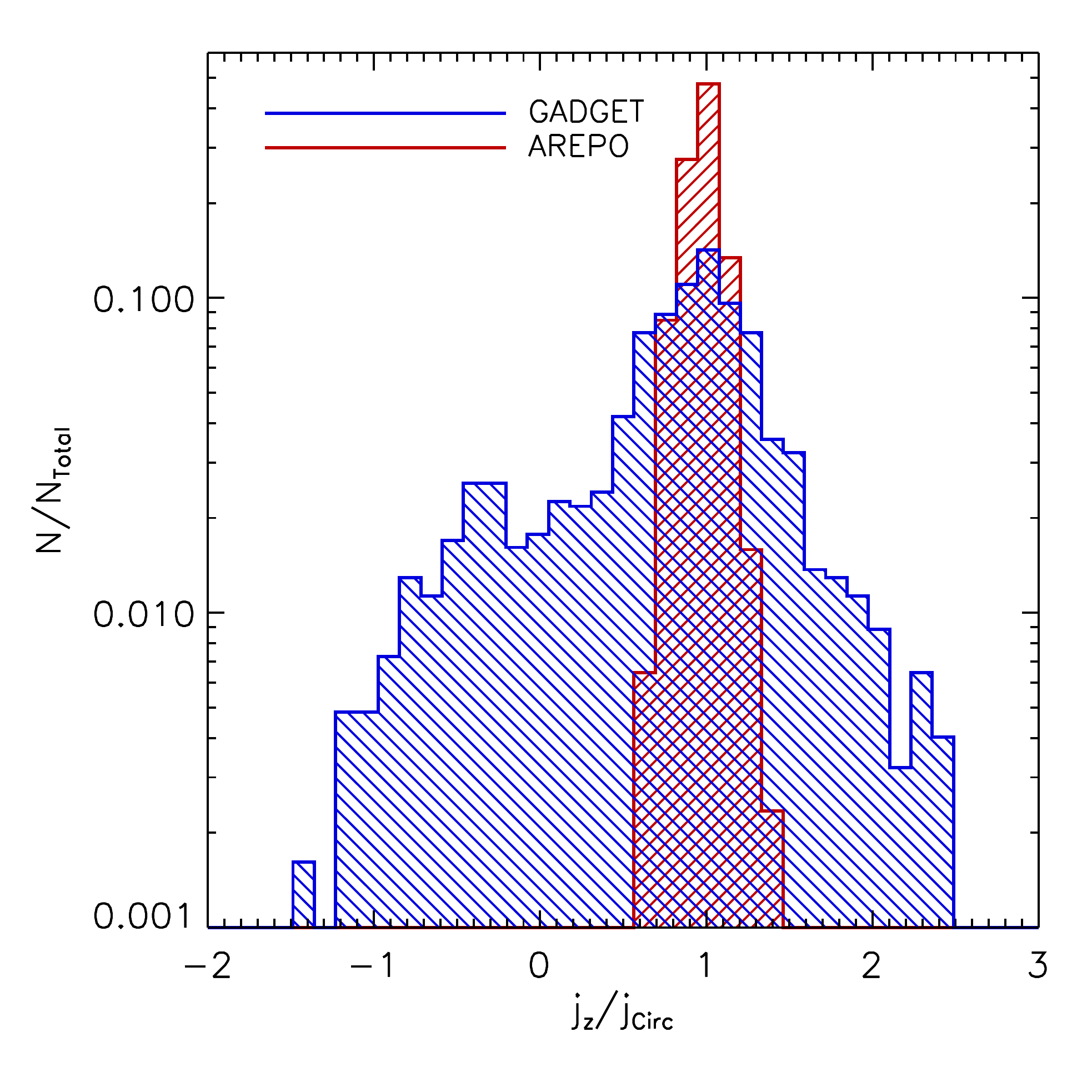}}}
\caption{The distribution of specific angular momenta in the disk for a
  typical matched galaxy in a $M_{ {\rm Halo} } = 10^{12} M_\odot$ halo.
  On the $x$-axis the
  ratio of the $\hat z$-component of the specific angular momentum to the
  expected specific angular momentum for a particle at that location on a
  circular orbit is shown. Both codes exhibit distributions that peak about $1$,
  which corresponds to a rotationally supported gas disk.  However, while the
  {\small AREPO} disk shows a narrow distribution, with most gas being on
  nearly circular orbits, the {\small GADGET} object has a much wider
  distribution, with some gas on highly non-circular trajectories.}
\label{fig:jz_over_jcirc}
\end{figure}

Differences between the {\small GADGET} and {\small AREPO} galaxies can
  also be seen via a histogram of the gas specific angular momentum, as shown
  in Figure~\ref{fig:jz_over_jcirc}. In this plot, where the specific
  angular momentum distribution for the example galaxies illustrated in
  Figure~\ref{fig:velocity_maps} is shown, there is a clear distinction
  between the two codes. The {\small AREPO}  galaxy shows a narrow
  distribution around $j_{circ}$ with most of the gas on circular orbits, and
  thus rotationally supported. The {\small GADGET} galaxy instead exhibits a
  much wider distribution around $j_z/j_{circ} \,= \, 1$ due to the significant gas
  component which is on highly non-circular orbits, with some material
  even counter-rotating, as evidenced by negative $j_z/j_{circ}$ values.

Figure~\ref{fig:AngMom} shows the specific angular momenta of the gas
disks and of all baryons as a function of galaxy baryon mass. The hatched
regions indicate where actual spiral and elliptical galaxies are
located, as noted by \citet{Fall1983}. We have verified that the
recent observational sample of~\citet{Courteau2007} falls within the
hatched region occupied by spirals. We find the specific angular
momenta of the cold gas and stellar material in the {\small AREPO}
simulation to be significantly larger than that of the {\small GADGET}
objects.  The larger specific angular momenta of the {\small AREPO}
galaxies indicates that they are more rotationally supported, which
accounts for their larger disk scale lengths.

\section{Origin of the Discrepancies}

In this paper we have presented a comparison of gas disks formed in
cosmological simulations \citep[][]{Vogelsberger2011} performed with the 
SPH based code {\small GADGET} and the moving-mesh code {\small AREPO}.  Both
codes use an identical gravity solver and include the same physical
processes (e.g., cooling, sub-grid model for star formation 
and feedback), but use fundamentally different hydro solvers.  Whereas
{\small GADGET} uses an SPH approach to evolve the gas, {\small AREPO}
uses a finite volume scheme on a moving Voronoi mesh.

Our primary conclusion is that the cold gas disks that form with
{\small AREPO} are described by notably different surface density
profiles than disks formed using {\small GADGET}.  We showed that, on
average, the cold gas disks in {\small AREPO} simulations have
significantly larger scale lengths compared to a matched sample of
{\small GADGET} disks.  Consistently, we also find higher specific
angular momenta for the {\small AREPO} disks.  Now that we have identified 
systematic differences in the disk scale lengths and angular 
momenta we discuss their principle numerical origins and address 
ways in which the discrepancies may be reduced.

\subsection{Spurious Hydrodynamical Torques}

It is well-known that conventional formulations of SPH suffer from
artificial angular momentum transport at phase boundaries -- like the hot halo
cold-disk transition.  For example, \citet{Okamoto2003} showed that SPH
simulations are prone to angular momentum transfer at this interface and that
this can have implications for disk formation in cosmological simulations.
Moreover, \citet{Okamoto2003} demonstrated that shearing flows at phase
boundaries are more accurately captured in grid-based hydro
solvers.  Thus, this is one particular area where we expect that the hydro
solver included in {\small AREPO} should yield more accurate results compared
to {\small GADGET}.

One solution to this problem is to completely decouple the ``hot" and ``cold"
particle neighbor searches~\citep{Okamoto2003,MarriWhite2002}.  In principle,
this modification does a better job at estimating the local gas density
using only particles that are in the same phase allowing for a cleaner
separation of multi-phase gas boundaries.  Since this will eliminate all
hydrodynamical interactions between the hot and cold phase the spurious  loss
of angular momentum will be eliminated.  However, decoupling the neighbor
searches for particles based on their phase could also lead to artificial
suppression of other physical phenomena that rely on direct interaction of
multi phase gas such as ram pressure stripping or shock capturing~\citep[see,
  e.g., the discussion in][for their procedure to address this 
  issue]{MarriWhite2002}.

Another solution proposed is to increase the simulation resolution
substantially~\citep{Okamoto2003,Kaufmann2007}.  Increasing the resolution
will reduce the influence of the pressure gradient mis-estimation that occurs at density
phase boundaries in standard SPH ~\citep[see e.g.][]{Agertz2007}. \citet{Kaufmann2007} tested the impact
of resolution on idealized inside-out disk formation simulations using SPH
and found that by increasing the number of SPH particles in a halo
above $10^6$ the artificial hydrodynamical angular momentum loss became
subdominant to other torquing mechanisms.  This problem is less severe (if
present at all) in grid based codes like {\small AREPO} where both density
phase boundaries and shearing flows are more accurately
handled~\citep{Okamoto2003, Agertz2007,Sijacki2011}.  As a result, simulated gas disks in {\small 
GADGET} rapidly loose their angular momentum unless they have a very large 
number of resolution elements while the same gas disks are able to evolve 
without such severe angular momentum loss at the same resolution in {\small 
  AREPO}.

\begin{figure}\centerline{
\includegraphics[width=7.9truecm,height=7.9truecm]{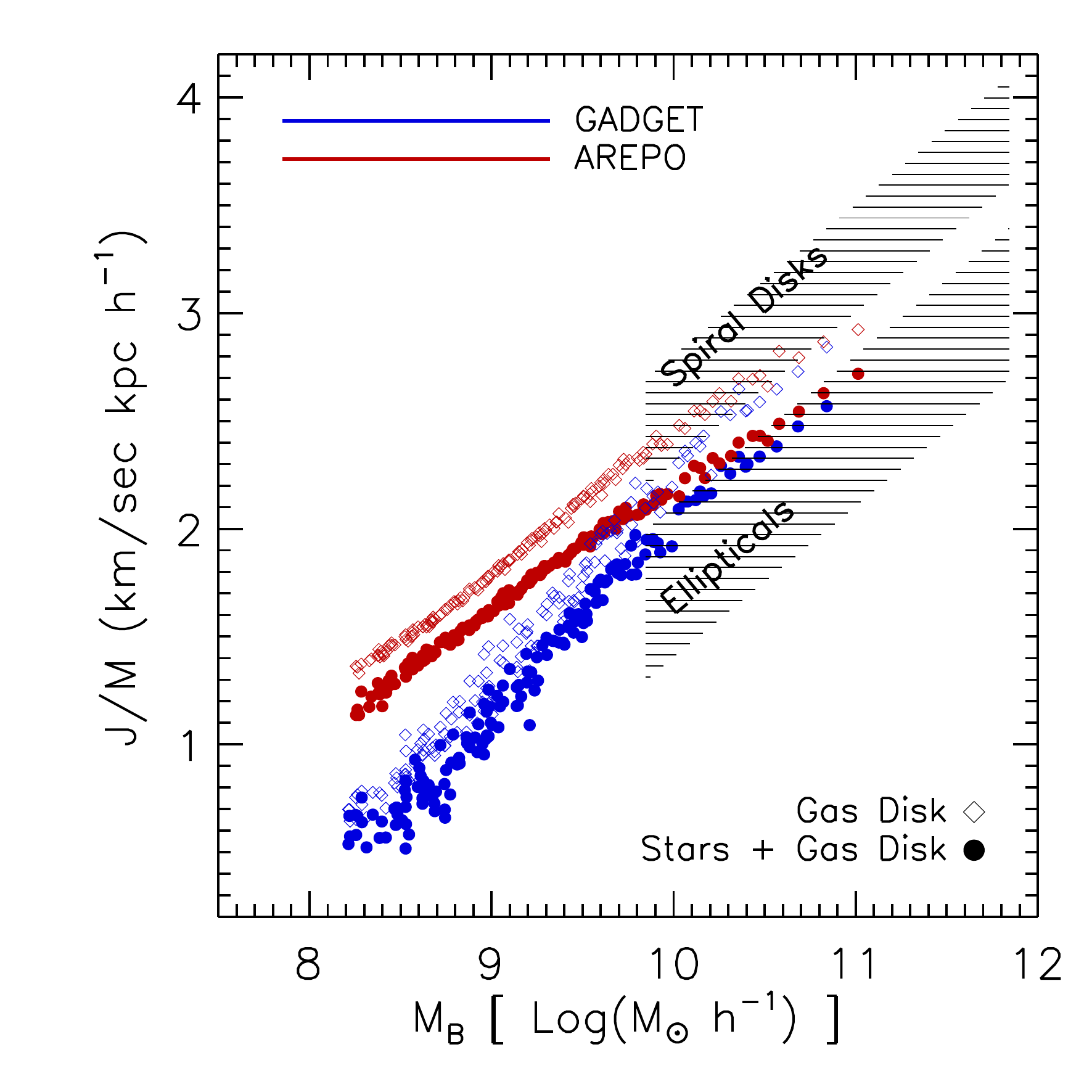}}
\caption{Specific angular momentum as a function of baryon mass
  for the gas (diamonds) and baryons (filled circles) for the matched
  sample of {\small AREPO} and {\small GADGET} galaxies.  Hatched regions
  denote locations of spirals and ellipticals on this diagram as defined by
  \citet{Fall1983}.}
\label{fig:AngMom}
\end{figure}

Since this artificial angular momentum transport is most prominent at low
resolution, this is the main reason for the differences seen between the disk
scale lengths of {\small GADGET} and {\small AREPO} objects at relatively low
masses (i.e. $< 10^{11} h^{-1}\; M_\odot $) and will contribute to the loss of angular
momentum for higher  mass systems (i.e. $< 10^{12} h^{-1}\; M_\odot $).  We note
that the most massive haloes considered in our current paper approach the
resolution criteria set forth by~\citet{Kaufmann2007} (i.e. $\sim 10^6$
particles per halo), so we do not expect spurious hydrodynamical angular
momentum loss to be the dominant problem in our high mass simulated
objects.  However, for the low mass objects in our simulation which are by
definition more poorly resolved, spurious angular momentum loss is bound to
have a substantially larger impact.  This conclusion is supported by 
examining the specific angular 
momentum content of the {\small GADGET} and {\small AREPO} galaxies -- 
as shown in Figure~\ref{fig:AngMom} -- which demonstrates that the 
discrepancy between the two codes is larger for low mass galaxies.

The resolution dependence of spurious hydrodynamical angular momentum transport 
demonstrated in~\citet{Kaufmann2007} implies that increasing the resolution of our 
{\small GADGET} simulations by an appropriate factor (an increase in the number of SPH particles 
of $10^3$ would give us the desired $10^6$ particles per halo at the low mass end) 
could improve the agreement between {\small GADGET} and 
{\small AREPO}.   In fact, since massive galaxies are 
assembled via hierarchically merging smaller objects together, it is necessary to have
$>10^6$ SPH particles in all haloes -- including low mass systems -- to avoid spuriously loosing 
angular momentum in low mass systems that will ultimately become part of a 
well resolved, high mass galaxies~\citep{Kaufmann2007}.  This is seemingly in accord with the results of very high
resolution ``zoom-in'' simulations~\citep[e.g.][]{Governato2004} that have found the 
angular momentum loss in gas disks formed in standard SPH simulations can be substantially 
reduced by increasing the particle resolution. However, while the $10^6$ SPH resolution elements 
per halo is a reasonable requirement for ``zoom-in'' simulations, this same requirement 
is not yet feasible for intermediate and low mass haloes in full cosmological box simulations.
In that sense, we consider it an advantage that grid based codes such as {\small AREPO}
do not suffer from this spurious hydrodynamical angular momentum loss even at
resolutions well below $10^6$ particles per halo.

\subsection{Gas Heating and Cooling}
Although {\small GADGET} and {\small AREPO} share the same prescriptions for
radiative gas cooling, non-adiabatic heat sources and mixing at phase
boundaries can cause differences in the growth of galactic gas
disks.  \citet{Vogelsberger2011} presented an analysis of the evolving
thermodynamic gas properties in our cosmological simulations.  One conclusion
from this analysis is that the cooling in {\small AREPO} is more efficient
than in {\small GADGET}, which is driven primarily by differences in the
cooling rates of diffuse halo gas.  There are two main reasons for the
differences in these cooling rates which are discussed in detail in
\citet{Vogelsberger2011} and summarized here.

The first reason is differences in dissipative heating in haloes driven by
the presence of turbulent energy.  \citet{BauerSpringel2011} performed a
comparison of the properties of simulated driven turbulence using {\small
GADGET} and {\small AREPO} and found that while the two codes produce
similar velocity and density power spectra for high mach numbers (i.e. for
highly supersonic driven turbulence), there are prominent differences  in the
way turbulent power cascades to smaller scales in the subsonic
regime.  In~{\small AREPO}, a Kolmogorov-like turbulent cascade is recovered
which transports energy to smaller spatial scales.  However, in~{\small
GADGET}, the turbulent large scale eddies are quickly dissipated close to 
the cooling radius and transformed into incoherent small-scale velocity noise
which is converted into thermal energy as the velocity noise is damped out
via artificial viscosity.  This heats gas and inhibits cooling in {\small 
  GADGET} haloes, driving part of the difference in the cooling rates seen 
in~\citet{Vogelsberger2011}.

The second reason stems from differences in mixing between the two codes,
especially at density phase boundaries.  For low mass galaxies, the gas cooling
timescale in the halo is relatively short, such that the material is able to
cool onto the central galaxy with similar efficiency in both codes.
However, for a typical massive galaxy in our simulations, the diffuse halo
gas becomes sufficiently hot and the gas cannot cool rapidly due to radiative
losses.  However, mixing that occurs around in-falling substructures as cool
gas is hydrodynamically stripped can substantially lower the local cooling
timescale~\citep{Marinacci2010}.  Idealized tests of gas stripping~\citep{Agertz2007,Sijacki2011}
show that this mixing will be suppressed in {\small GADGET}.  

A similar mixing boundary layer can develop at the interface of the  central
gas disk and the diffuse hot halo.  This shearing phase boundary can generate
mixing via Kelvin-Helmholtz instabilities which will be poorly resolved in
{\small GADGET}~\citep{Okamoto2003}.  Although the diffuse halo gas sitting
just above the gas disk boundary may have long cooling times, continued
efficient hot-mode gas accretion can be facilitated by mixing at this
boundary.  \citet{Sijacki2011} demonstrated this point by examining the
cooling rates of gas in an idealized gas sphere with {\small GADGET} and
{\small AREPO}.  They find that the cooling rates are nearly identical when no
gas rotation is included, which validates that the cooling prescriptions in
both codes are in fact functionally identical.  However, when the same test is
repeated with gas rotation the cooling rates for the two codes became
discrepant, with {\small AREPO} cooling more efficiently.  The more efficient
cooling only sets in once a substantial amount of gas settles into a
rotationally supported disk that can interact with the ambient halo gas.
\citet{Keres2011} present evidence for enhanced cooling of this same sort in
massive galaxies in the cosmological simulations by noting that the hot haloes
in {\small AREPO} have cooler cores compared to {\small GADGET} and the gas
in {\small AREPO} tends to be on radially inward trajectories -- both of
which are consistent with a scenario of less subsonic turbulence dissipation and of 
increased cooling in a mixing boundary
layer between the diffuse halo gas and dense gas disk.

The combined effect of ``extra heating'' in {\small GADGET} from the poorly
resolved turbulent power cascade and ``extra cooling'' in {\small AREPO} from
better resolution of mixing at phase boundaries explain the global
thermodynamic differences seen in~\citet{Vogelsberger2011}.  This also
explains part of the differences in disk scale lengths that we see in this
paper.  Notably, high mass galaxies in {\small AREPO} will be more efficient
at accreting material from the hot halo at late times which will help them
maintain a gas rich disk.  

It is worth noting that several proposals have been put forward to improve the
mixing at phase boundaries in SPH, which may bring the two codes into better
agreement.  For example, it has been suggested that by including a thermal
conductivity term contact discontinuities and instabilities can be more
accurately handled ~\citep[e.g.,][]{PriceThermalConductivity,
WadsleyMixing2008}, which would improve the mixing picture with respect to
the standard SPH.  Alternatively, moving away from the density based
formulation of SPH to an energy or pressure formulation can substantially
reduce the artificial surface tension at contact discontinuities, which can
help to resolve instabilities and mixing with higher 
accuracy~\citep[e.g.,][]{RitchieThomas2001,SaitohMakino2012}.   In particular,
\citet{Hopkins2012} has recently shown how such a formalism can be
derived from a variational principle, as in~\citet{SHentropy}, resulting
in a fully conservative version of SPH.  Other changes,
such as modifying the shape of the adopted smoothing kernel with a substantial
increase in the number of neighbors used in the hydro calculations, or
modifying the momentum equation evaluated in the code, have all also shown
promise in improving the performance of SPH at resolving
instabilities~\citep[e.g.,][]{rpSPH,SPHS}.  


\begin{figure*}
\centerline{\vbox{\hbox{
\includegraphics[width=7.9truecm,height=7.9truecm]{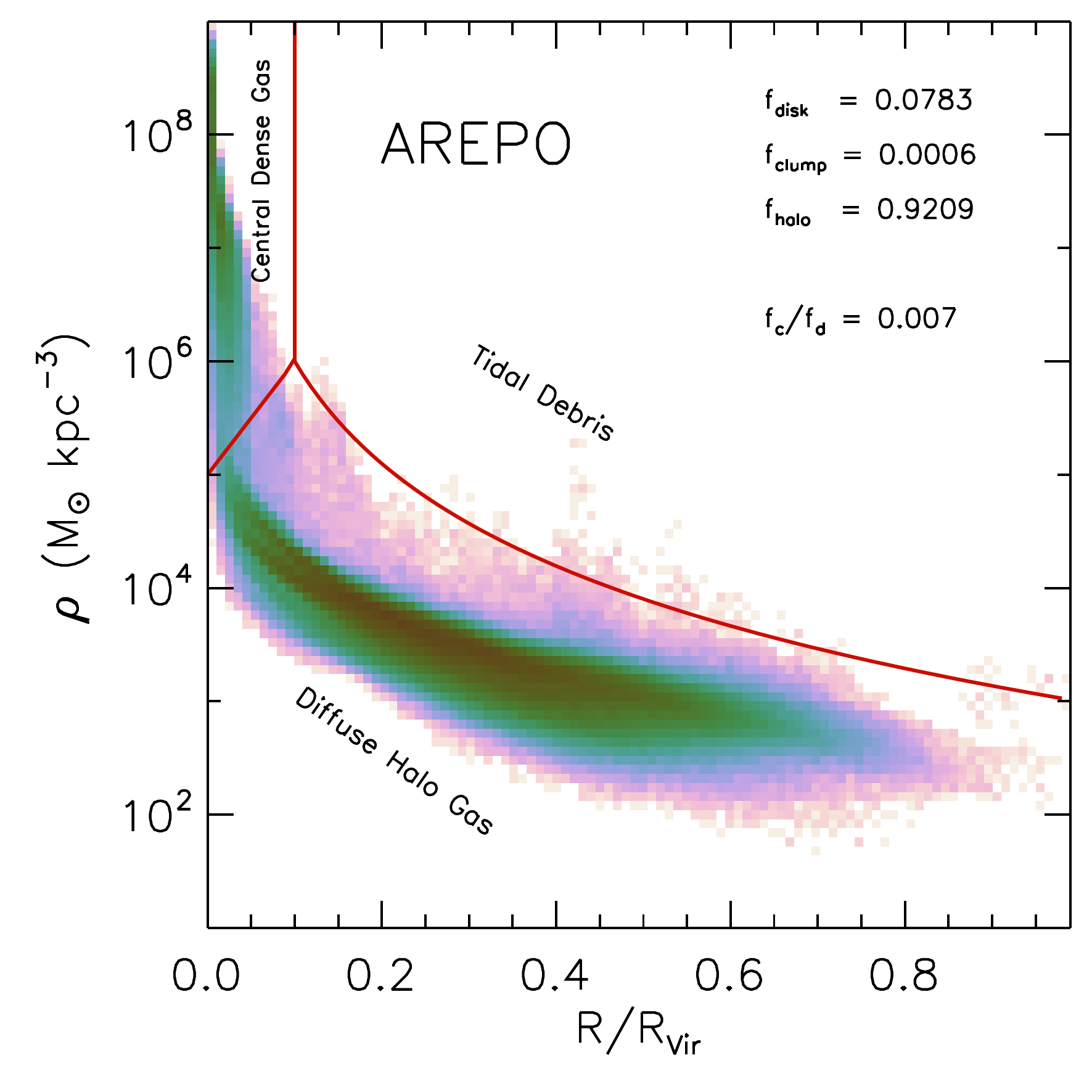}
\includegraphics[width=7.9truecm,height=7.9truecm]{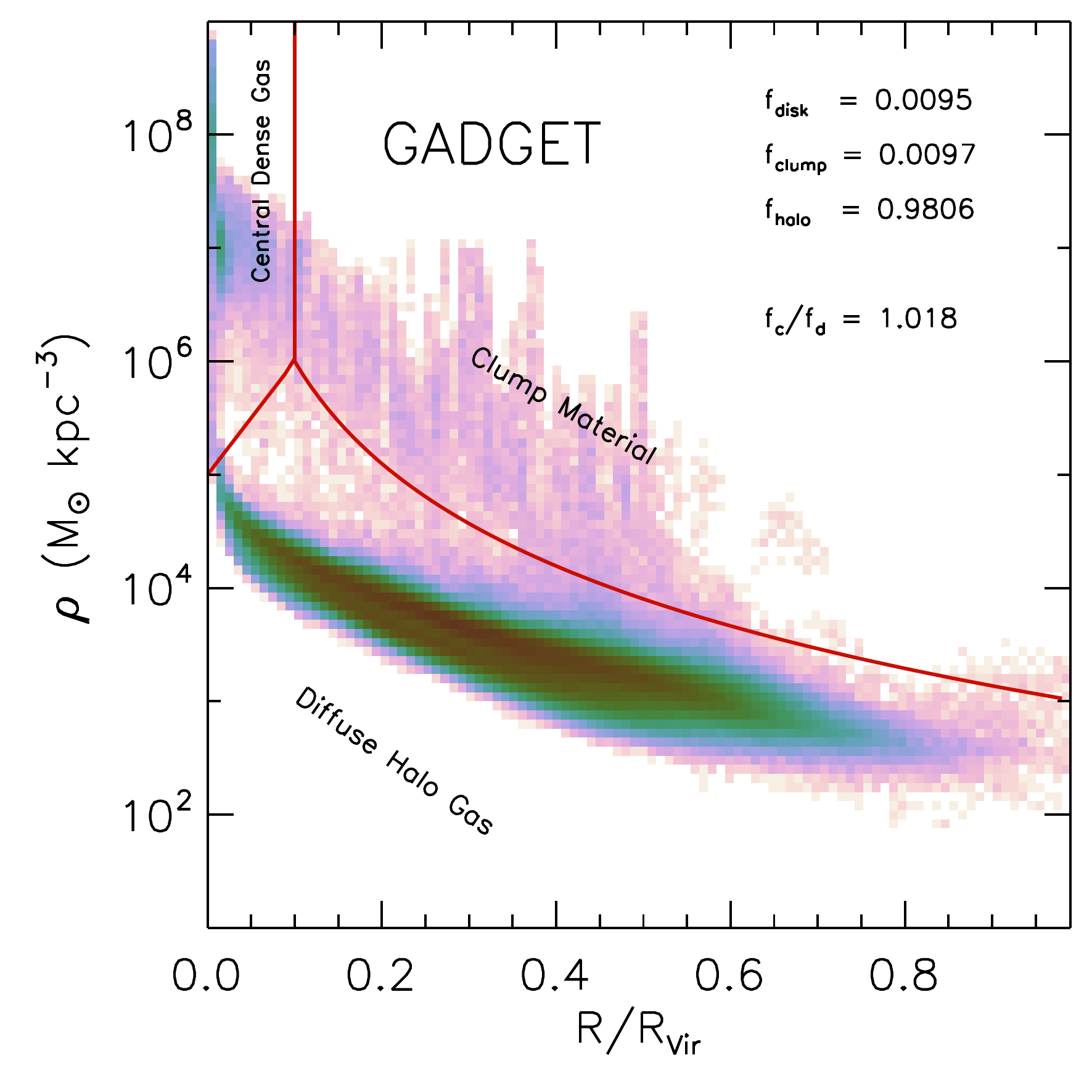}}
}} 
\caption{The distribution of gas in the $5$ matched  haloes between $10^{12}$
  and  $3\times10^{12} h^{-1}\; {\rm M}_\odot$ in {\small AREPO} (left) and
  {\small GADGET} (right). Red lines mark the boundaries  we have chosen to
  separate the diffuse halo gas, central dense gas, and clump material are
  described in detail in the text. The fraction of gas that resides in the
  disk, $f_{\rm disk}$, halo, $f_{\rm halo}$, and clump, $f_{\rm clump}$
  regions are reported on each panel. Finally, the ratio of the clump gas
  mass to the disk gas mass, $f_{\rm c}/f_{\rm d}$ is also shown.}
\label{fig:r_versus_rho}
\end{figure*}

\subsection{Cold Clumps}
\label{sec:ColdClumps}
One important issue contributing to the differences in the disk scale  lengths
for high mass objects in {\small GADGET} and {\small AREPO} is the efficient
accretion of low angular momentum gas via dense clumps. These clumps can be
quite easily identified using projected surface density maps of the gas
around relatively massive galaxies residing in $\sim 10^{12} {\rm M_{\odot}}$
haloes in {\small GADGET} as is shown in
Figure~\ref{fig:PostageStamps_z0_blobs} and~\ref{fig:PostageStamps_z2_blobs}
at redshift $z=0$ and $z=2$, respectively. In contrast, very few blobs are
present in the {\small AREPO} galaxies aside from discrete objects which we
have identified to be subhaloes with clearly associated dark matter
components.

\subsubsection{Identifying Clump Material}
We can identify the clump  particles by noting that they are overdense
relative  to the ambient hot halo density, colder than the ambient  hot halo
temperature, and do not have any substantial  dark matter overdensity
associated with them. To demonstrate  this point,
Figure~\ref{fig:r_versus_rho} shows a 2-D histogram of the material that is
part of the primary subhalo (as identified via {\small SUBFIND}) in the $5$
matched haloes between  $10^{12} h^{-1} \; {\rm M}_\odot$ and 
$3 \times 10^{12} h^{-1} \; {\rm
  M}_\odot$ at redshift $z=0$ in density-radius space (also shown in
Figure~\ref{fig:PostageStamps_z0_blobs}). Although the clumps can be as or
even more prominent at higher redshifts, here we focus on this mass range and
redshift in our analysis here for demonstrative purposes.

We can place rough boundaries to break the density-radius space  down into
three physically meaningful  components: central dense gas, diffuse hot halo
gas,  and cold/dense blobs. The central dense gas -- which  has been the
subject of most of this paper -- is concentrated into a  relatively small
region at small radii and has high density
\begin{equation}
M_{ {\rm disk}} = M \left( \frac{r}{R_{ {\rm vir}} } < 0.1 , \rho > 10^{\frac{5 + r }{ (0.1 R_{{\rm vir}})}} \left[\frac{ {\rm M}_\odot h^2 }{ {\rm kpc}^{3} } \right]  \right).
\end{equation}
The dense clumps are identified as dense material outside of the noted
central dense gas region, i.e.,
\begin{equation}
M_{ {\rm clump}} = M \left( \frac{r}{R_{ {\rm vir}} } > 0.1 , \rho >  \frac{10^3}{ (r/R_{{\rm vir}})^3} \left[ \frac{ {\rm M}_\odot h^2 }{{\rm kpc}^{3} } \right] \right).
\end{equation}
Finally, we assign all other material to be part of the diffuse hot halo which
is characterized by relatively low density.

The fraction of mass that resides in each region is noted  on both panels of
Figure~\ref{fig:r_versus_rho}. In both {\small GADGET} and  {\small AREPO}
over $90$\% of the gas mass associated with these  systems resides in the
diffuse  halo region. \citet{Vogelsberger2011} and~\citet{Keres2011} analyzed
the hot halo gas and found that  more hot halo gas is present and  that the
temperature of this gas is hotter in {\small GADGET} compared to {\small
  AREPO}. This is consistent with our discussion from the previous  section on
the increased artificial heating in {\small GADGET} and increased  mixing
induced cooling in {\small AREPO}.

\begin{figure}\centerline{\vbox{\hbox{
\includegraphics[width=7.9truecm]{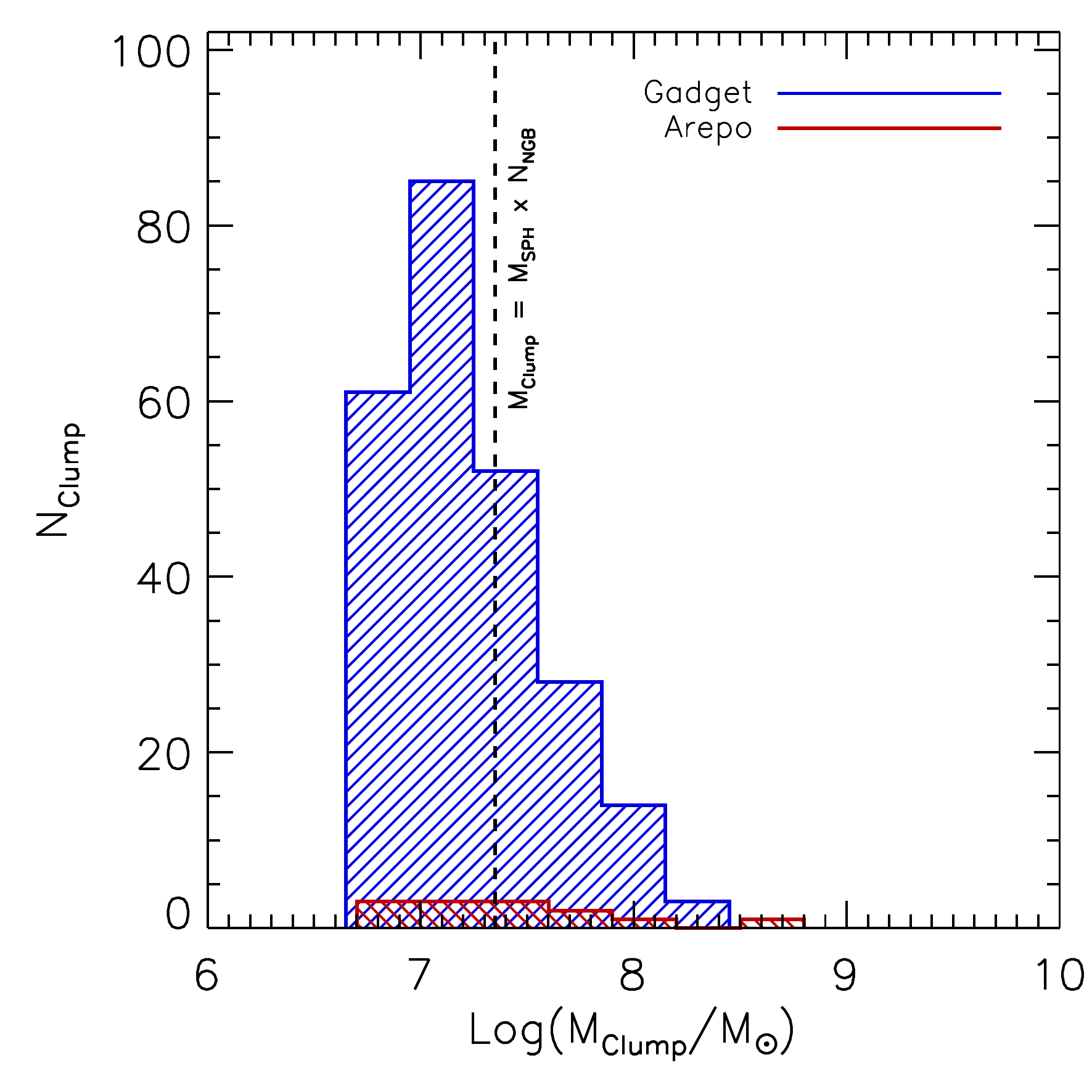}}
}}
\caption{  The distribution of clump masses is shown for {\small GADGET} (blue)
  and {\small AREPO} (red).  We find that there is a substantially larger
  population of clumps in the {\small GADGET} simulation, and that these
  clumps build up around the mass resolution limit of our simulation which is
  marked by the vertical dashed line.  We emphasize that the few 
  ``clumps'' found in the {\small AREPO} simulation are actually 
  tidal features or recently stripped cold gas, 
  which are identified as clump material according to our imposed 
  density threshold.  The overwhelming majority of the {\small GADGET} 
  gas clumps do not contain dark matter and are 
  not associated with infalling substructure or tidal features, 
  as can be gleaned from 
  Figures~\ref{fig:GadgetCenterpiece} and~\ref{fig:PostageStamps_z0_blobs}.    }
\label{fig:ClumpProperties2}
\end{figure}

The central dense region in {\small AREPO} contains about $8$\% of the total
gas mass, compared to about $1$\%  in {\small GADGET}. We identify two simple
reason for this difference. First, more gas is able to cool into this region
in {\small AREPO} for the reasons discussed in the previous section. Second,
we find that the {\small AREPO} galaxies are able to maintain larger amounts
of gas in this region because the dense gas resides in a rotationally
supported  disk with intermediate star formation rates. In contrast, we find
that the {\small GADGET} galaxies  in these massive systems contain most of
their gas mass  in a very centrally concentrated region with high densities,
which efficiently convert the gas supply into stars.

We now turn to the ``clump material'' portion of this diagram.  As we
described earlier, material that resides in this region is far away from the
central galaxy, but very dense compared to the ambient halo
gas. Interestingly, we find a substantial population of cold clump gas in
the {\small GADGET} objects that is not  found in the {\small AREPO}
systems. In terms of the fractional mass distribution,  $\sim 1$\% of the gas
mass is in this region for the {\small GADGET} systems, while a negligible
fraction is in the same region for {\small AREPO} objects.  Furthermore, 
of the small gas fraction that does reside 
in this region for the {\small AREPO} objects, most of this material 
is associated with tidal features or recently stripped cold gas from infalling satellites.

To quickly
estimate the potential impact of these clumps on the growth of the  central
disk, we report the ratio of clump mass to disk mass in
Figure~\ref{fig:r_versus_rho}. In {\small GADGET} we find that a comparable
amount of mass is in the central dense region as in the clump region.  In
other words, if these clumps are able to efficiently migrate  toward the
central object -- which we  will soon argue is the case -- then they are
capable of contributing a substantial amount of  cold, low-angular momentum
material to the  central galaxy. We note that although these quoted fractional
mass distributions depend on our choice for the boundary  locations shown in
Figure~\ref{fig:r_versus_rho}, our conclusion that there is a substantial
amount of mass in the clump region for the {\small GADGET} objects in this
mass range is robust to any reasonable changes in the boundary definition.

\begin{figure*}
\centerline{\vbox{\hbox{
\includegraphics[width=7.9truecm,height=7.9truecm]{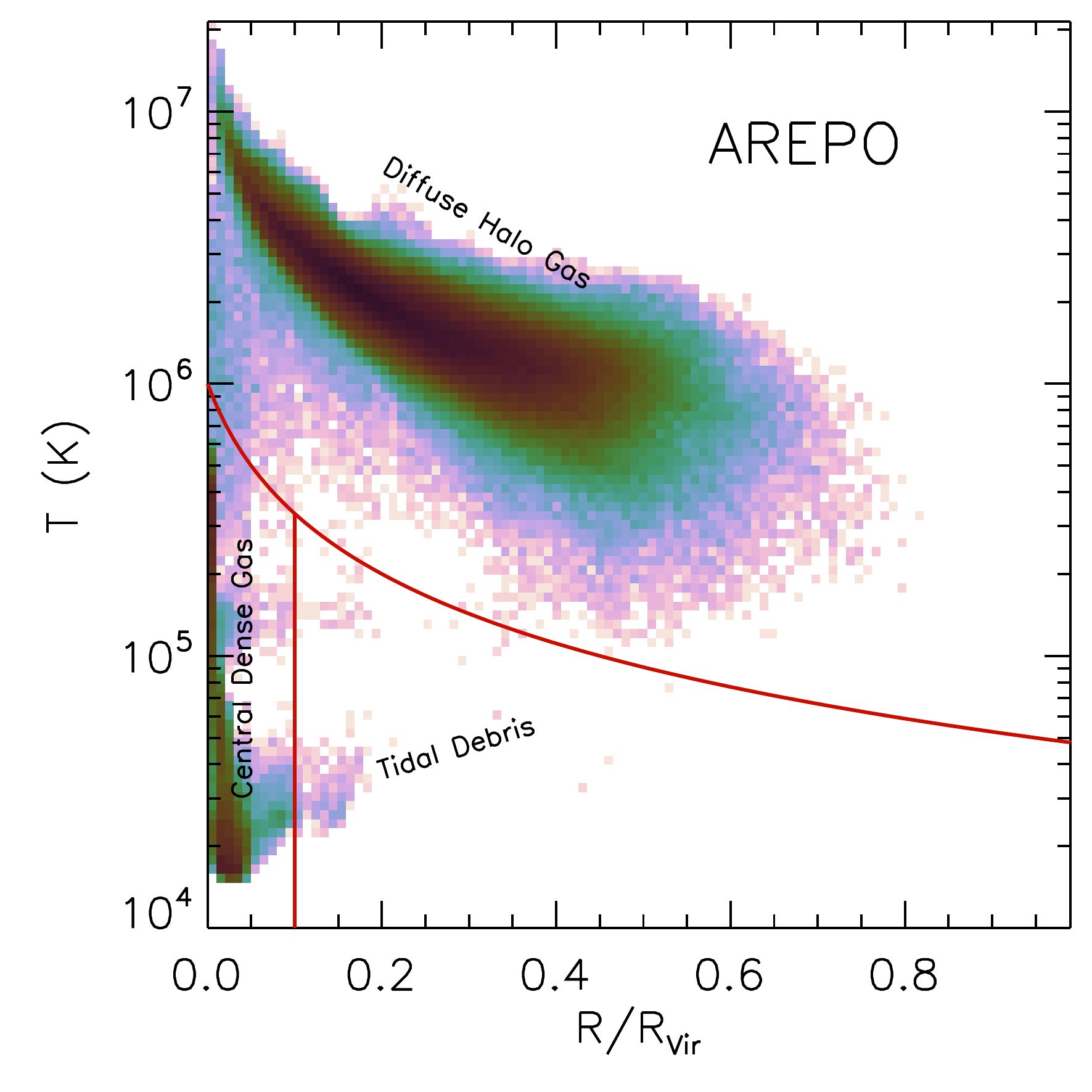}
\includegraphics[width=7.9truecm,height=7.9truecm]{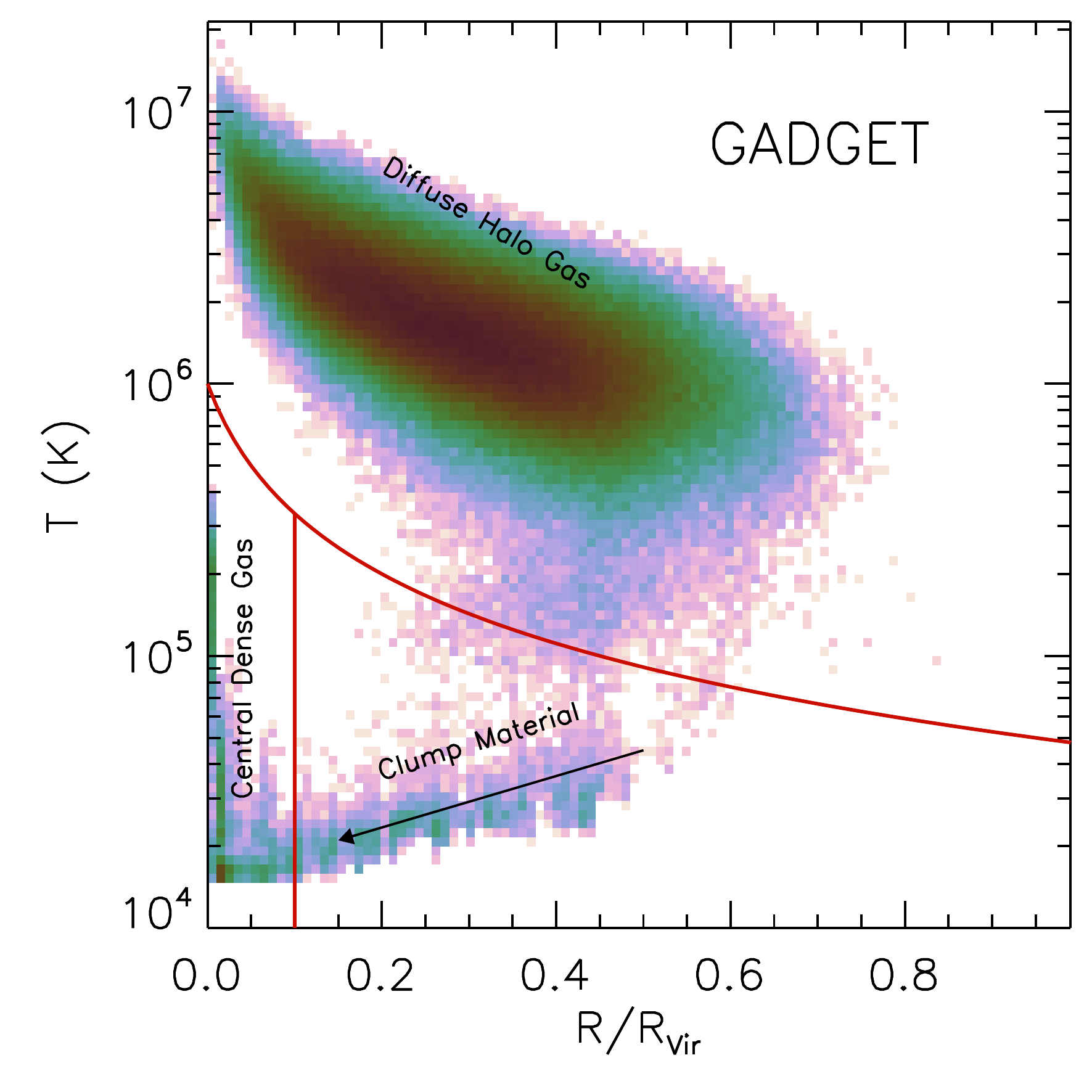}
}}}

\caption{Temperature distribution of bound gas in the $5$ matched  haloes
  between $10^{12}$ and  $3\times10^{12} h^{-1} {\rm M}_\odot$ in  {\small
    AREPO} (left) and {\small GADGET} (right). Red lines mark the boundaries
  we have chosen to separate the diffuse  halo gas, central dense gas, and
  clump material.}
\label{fig:rt}
\end{figure*}

\subsubsection{Clump Associations}

To classify these clumps, we select all overdense particles 
according to the cut presented in Figure~\ref{fig:r_versus_rho} and group
these particles using a Friends-of-Friends (FOF) 
algorithm with a linking length of $l=3 h^{-1}\;{\rm kpc}$ yielding a  
set of particles belonging to each clump. Each clump then has a well defined 
position, velocity, and mass.  

One quantity of interest is the dark matter overdensity associated with these
clumps. Since we have included only material that is bound to the primary
halo within our FOF group, we expect that well defined  substructure (e.g.,
in-falling dwarf galaxies) will have already been removed. We can verify that
this is the case by checking the dark matter overdensity  associated with each
clump, and comparing it to the spherically averaged dark matter density at
the clump's position.  In practice, we do this by finding the volume
associated with the $N^{th}$ nearest dark matter particle from the clump's
center of mass, and compare that to the spherically averaged dark matter
density measured in a thin shell at the same radius  as of the clump. Although
we have verified that this method would allow us to clearly identify
substructure, we do not find any clumps in our sample with significant
associated dark  matter overdensities. We have repeated our overdensity test
using $N^{th}={64,100,500}$ without any change in our results.


The mass spectrum for these blobs is  shown in
Figure~\ref{fig:ClumpProperties2}. We find that the majority  of the blobs
contain just above or below 32 particles, which corresponds to the number of
nearest neighbors used in our SPH simulation as denoted by the vertical
dashed line in Figure~\ref{fig:ClumpProperties2}. Clumps tend to build up
around the resolution limit of our simulation and as we have verified the mass
spectrum changes accordingly if we increase or decrease the number of nearest
neighbors.   Although we do find a very small number of low mass cold and dense gas patches 
in the {\small AREPO} simulation, we note that these features are characteristically 
distinct from the cold gas clumps in {\small GADGET}.  Specifically, these cold 
gas patches tend to be associated with tidal features or recently stripped 
gas from infalling satellites.  So, while these cold gas patches meet the density based selection 
criteria that we have implemented and do not contain any clearly associated dark 
matter overdensity, we emphasize that their origin is very different from the 
large population of cold gas clumps that are seen the {\small GADGET} simulation.

\begin{figure*}\centerline{\vbox{\hbox{
\includegraphics[width=7.9in]{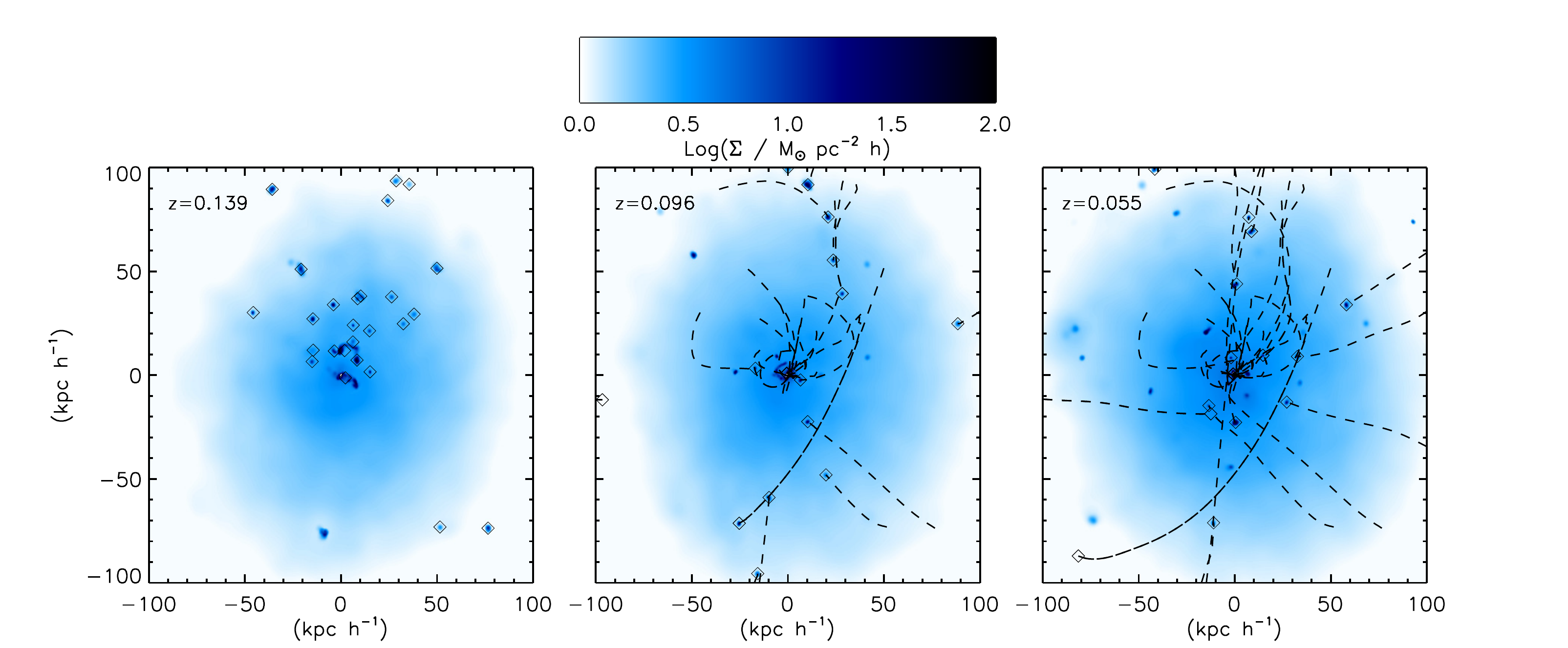}}
}}
\caption{  Maps of the gas surface density showing the trajectories of cold, dense gas clumps for a $\sim 10^{12}  h^{-1} \; M_\odot$ halo in the {\small GADGET} simulation at redshifts $z=\{0.130, 0.096, 0.055\}$ (left, central and right panel, respectively).  A population of clumps (marked with diamond symbols) are identified in the leftmost panel and tracked forward in time.  Dashed lines are denoting their trajectories. The vast majority of the clumps shown are moving toward the central galaxy on nearly radial trajectories, and are able to merge with the central galaxy. Thus, the clumps effectively provide a source of cold, low angular momentum material in the innermost regions, which is of entirely numerical origin. }
\label{fig:ClumpTrajectories}
\end{figure*}

\subsubsection{Clump Origin and Trajectory}

The origin of these clumps can be determined by tracing clump particles back 
in time.  We find that the clump particles originate in very mild gas
over-densities within the filamentary structures of the intergalactic medium
(IGM). As  material in these filaments falls into the hot halo  environment
that surrounds central galaxies  in massive objects, mild over-densities are
amplified via hydrostatic pressure confinement.  We find that the maximum
past temperature of the clumps is well below the  halo virial temperature,
indicating that the blobs did not form via cooling instability of gas
overdensities present in the hot halo~\citep[as studied in,
 e.g.,][]{Kaufmann2006}, but rather remained cold during their accretion from
 the IGM.  This point is demonstrated in Figure~\ref{fig:rt} which shows a 2-D
 histogram of the distribution of gas in radius-temperature space for $5$
 matched objects with halo masses just above $M=10^{12} h^{-1} \; M_\odot$ at
 redshift $z=0$.  As labeled within the plot the trajectory for
 clumps can be readily identified for the {\small GADGET} haloes, which
 allows cold material to migrate from large radii to the central object
 without ever heating substantially.  No analogous migration trajectory exists
 for the {\small AREPO} systems. The clump formation picture we have discussed
 here is consistent with the clump formation scenario outlined
 in~\citet{Keres2009}, where an analogous population of cold gas clumps were
 found originating from IGM filaments.  Also in agreement
 with~\citet{Keres2009}, we find an accretion rate from these cold clumps in 
 our {\small GADGET} simulation of $\dot M _{{\rm gas}} \sim 0.5 M_\odot {\rm
 yr}^{-1}$ at redshift $z=0$ for $M=10^{12} h^{-1} \; M_\odot $ systems.

As a clump is gravitationally accelerated toward the halo center, the clump 
should begin to be disrupted and mixed via ram pressure stripping and the 
Kelvin-Helmholz and Rayleigh-Taylor instabilities.  However, it is well known 
that these instabilities are poorly resolved and that ram pressure stripping 
is underestimated in the standard density formulation of SPH.  In
particular,~\citet{Agertz2007} and~\citet{Sijacki2011} presented numerical 
experiments that showed cold blobs have artificially long survival times in 
SPH codes, while grid based codes like {\small AREPO} shred these clumps over
substantially shorter timescales consistent with analytic  expectations.  As a
result, mild over-densities in the accreting filamentary material are
condensed and fragmented into a  population of high density cold clumps that 
have substantially longer survival times in the {\small GADGET} simulation
compared to {\small AREPO}.  Since idealized numerical experiments have shown
that the survival timescale for cold gas blobs in {\small GADGET} is
artificially long compared to analytic expectations, we argue that the
survival of these blobs in our {\small GADGET} cosmological simulations is a
consequence of the same SPH deficiencies.

When the clumps enter the virial radius for the first time, they have non-zero orbital 
angular momenta about the halo's center of mass that is consistent with other
recently accreted material. However, as the clumps pass through the halo gas, 
they loose their angular momentum efficiently due to spuriously strong 
hydrodynamic drag forces~\citep{Tittley2001} and dynamical friction.  This 
loss of angular momentum puts the clumps on increasingly radial trajectories 
and allows them to merge with the central object after only one or two orbits.  Thus,
the clumps become an efficient source of cold, low angular momentum gas feeding 
the central galaxy.

Figure~\ref{fig:ClumpTrajectories} specifically highlights the trajectories of 
a population of clumps in time.  A set of cold, dense gas clumps are identified 
in the leftmost panel of Figure~\ref{fig:ClumpTrajectories} and their subsequent
trajectories are marked with dashed lines in the central 
and rightmost panels.  By inspection of the 
marked clump trajectories, it can be seen that many clumps move on
nearly radial trajectories and eventually merge with the central galaxy.  Although 
each clump has a relatively low mass ($\sim 10^7-10^8 h^{-1} \; M_\odot $ -- see 
Figure~\ref{fig:ClumpProperties2}), a sufficiently large 
number of clumps fall into the central galaxy on a characteristic time scale of $\sim 1 {\rm Gyr}$. Thus, their cumulative mass amounts to a substantial fraction of the central galaxy gaseous mass. Furthermore, given that they arrive on nearly 
radial trajectories with low angular momentum, the clumps act as an efficient delivery source of 
low angular momentum fuel to the central galaxy.

%


A similar population of blobs can be seen in the gas distribution 
for relatively massive haloes (i.e. $\sim 10^{12} h^{-1} \; M_\odot $) 
in independent studies that used similar versions of 
{\small GADGET}~\citep[e.g.,][]{vandeVoort2011,Keres2009letter,vandeVoort2012}.
This is an interesting point because the mass resolution used 
in~\citet{vandeVoort2011} is about a factor of $2$ worse than the resolution 
used in our present study, and the resolution  used in~\citet{Keres2009letter} 
is a factor of $7$ better.  Both of these studies find that these clumps 
result from fragmented IGM filaments that are able to survive until they 
merge with the central gas disk.  Despite their substantially higher 
resolution, \citet{Keres2009letter} find that these clumps tend to form just 
above the resolution limit of their simulation -- implying a significant 
change in the clump mass spectrum from what we have found here.  

 Recently, \citet{Hobbs2012} have made use of a new flavor of SPH~\citep{ReadHayfield2012}
to study the formation and impact of ``blobs'' in standard 
SPH simulations and found their origin to lie in artificial thermal instabilities that 
can originate from a small number of particles with low entropies with respect to their
neighbors.  Since the standard formulation of SPH in {\small GADGET} 
lacks any inter-particle fluid mixing, these artificial thermal instabilities are 
allowed to grow and ultimately form a population of dense gas clumps.  
\citet{Hobbs2012} show that this artificial thermal instability can be averted 
by including thermal conductivity~\citep[e.g.,][]{PriceThermalConductivity}.  
This issue does not arise with {\small AREPO} because cells exchange 
entropy with their neighbors when mass is advected across cell boundaries, 
which results in a physically motivated homogenization of the fluid entropy.
The presence of these clumps in independent numerical studies of 
varying resolution~\citep[e.g.][]{vandeVoort2011,Keres2009letter,vandeVoort2012} seems 
to imply that these clumps will continue to form and survive at higher 
resolution even though the specific properties, such as the mass spectrum, 
will change unless modifications are made to the standard SPH hydro solver.

\section{Conclusions}

Here we presented a comparison project aimed at studying the properties  of
gas disks formed in cosmological simulations performed with two  very
different hydrodynamical codes -- {\small GADGET} and {\small AREPO}.  Our
comparison  started from an identical set of initial conditions which were
evolved forward in time with the two codes which have the same gravity solver
and while holding fixed the initial number of resolution elements, and radiative
cooling, star formation and feedback prescriptions. However,
{\small GADGET} and {\small AREPO} adopt very different
approaches for solving the  hydrodynamic equations. While {\small GADGET} uses
a  standard density formulation of SPH, {\small AREPO} solves the fluid
equations using a Riemann solver on an unstructured moving mesh. In many ways,
the hydro solver included in {\small AREPO} has accuracy advantages over the
SPH solver used in {\small GADGET}  which can be clearly demonstrated in,
e.g., idealized shock tube tests,  driven turbulence tests, and hydro instability
tests. The goal of our comparison was to understand the impact of the
hydro solver on the formation of gas disks in fully cosmological simulations
at a comparable resolution. Our primary conclusions are as follows:

\begin{itemize}
\item  After fitting the gas disks with best-fit exponential surface 
density profiles, we find that the {\small AREPO} gas disks are systematically
larger than their {\small GADGET} counterparts. This corresponds to 
gas disks in {\small GADGET} having lower specific angular momentum compared 
to the matching set of disks formed in {\small AREPO} simulation.

\item The primary reason responsible for the differences in gas disk  scale
  lengths between the two codes changes as a function of the number of
  resolution elements and physical environment of the host halo.

\item For low mass objects, low resolution leads to spurious angular momentum
  transport from the cold disk to the diffuse hot halo in the {\small GADGET}
  simulation. This spurious angular momentum loss is a well-known and
  documented issue, which can be alleviated by moving to increasingly higher
  resolution in test problems or ``zoom-in'' simulations. However, for large
  cosmological box simulations, like those we have presented here, the
  resolution needed to suppress this spurious angular momentum loss is not yet
  attainable. Grid based  codes -- such as {\small AREPO} -- are not expected
  to suffer from this same problem and  can therefore provide a more accurate
  answer for the same number of resolution elements and comparable CPU time.

\item Poorly resolved subsonic turbulence in {\small GADGET} results in 
  dissipative heating of the gas near the cooling radius.  This inhibits 
  the accretion of gas onto the central galaxy.  In the presence of turbulent energy,
  {\small AREPO} correctly recovers a cascading Kolmogorov-like power spectrum,
  so no analogous artificial heating source is present.

\item For high mass objects, the cooling timescale of hot halo  gas can become
  comparable to the Hubble time which effectively  shuts off fresh gas
  accretion. However, mixing at density  phase boundaries -- such as the
  interface between the cold gas  disk and the hot halo -- can substantially
  increase the  gas cooling rates. This  allows for more gas to cool onto the
  disk in {\small AREPO} given that the mixing at phase boundaries is
  suppressed in {\small GADGET}.

\item For high mass objects, the efficient delivery of low angular momentum
  gas  in the form of cold gas clumps causes the central gas disks in {\small
    GADGET} to be much more centrally concentrated than in {\small
    AREPO}. These clumps form from fragmented IGM filaments and rapidly
  migrate to the potential minimum as they loose their angular momentum to
  hydrodynamic drag against the ambient hot halo. The absence of these blobs
  in {\small AREPO} is attributed to the efficient disruption of clumps via
  ram pressure stripping and the Kelvin-Helmholtz and Rayleigh-Taylor
  instabilities -- all of which are  poorly handled in {\small GADGET}.

\end{itemize}

\section*{Acknowledgements} 
We thank Elena D'Onghia for helpful suggestions on this
work and St\'ephane Courteau for providing observational data in tabulated
form. DS acknowledges NASA Hubble Fellowship through grant HST-HF-51282.01-A.
\bibliographystyle{mn2e}

\end{document}